\documentclass{aa}  

\usepackage{graphicx}

\usepackage{txfonts}
\usepackage{xcolor}
\usepackage{amsmath}
\usepackage{float}

\usepackage{array}
\usepackage{diagbox}
\usepackage{subcaption}

\usepackage[colorlinks=true, allcolors=blue]{hyperref}
\begin{document} 

    \title{BH-BH mergers with \& without EM counterpart}
   \subtitle{A model for stable tertiary mass transfer in hierarchical triple systems}
   
   \author{F. Kummer
          \inst{1}
          \and
          S. Toonen\inst{1}
          \and A. Dorozsmai\inst{2}
          \and E. Grishin\inst{3,4}
          \and A. de Koter\inst{1,5} 
          }

   \institute{Anton Pannekoek Institute for Astronomy, University of Amsterdam, Science Park 904, 1098 XH Amsterdam, Netherlands\\
              \email{f.a.kummer@uva.nl}
              \and 
              National Astronomical Observatory of Japan, National Institutes of Natural Sciences, 2-21-1 Osawa, Mitaka, Tokyo 181-8588, Japan
              \and
              School of Physics and Astronomy, Monash University, Clayton, VIC 3800, Australia
              \and
              OzGrav: Australian Research Council Centre of Excellence for Gravitational Wave Discovery, Clayton, VIC 3800, Australia
              \and
              Institute of Astronomy, KU Leuven, Celestijnenlaan 200D, 3001 Leuven, Belgium\\ 
             }

   \date{Received ... ; accepted ...}

 
  \abstract
   {Triple stars are prevalent within the population of observed stars. 
   Their evolution, compared to binary systems, is notably more complex, influenced by unique dynamical, tidal and mass transfer processes inherent to higher order multiples. Understanding these phenomena is essential for a comprehensive insight into multistar evolution and the formation of energetic transients, including gravitational-wave (GW) mergers.}
   {Our study aims to probe the evolution of triple star systems when the tertiary component fills its Roche lobe and transfers mass to the inner binary. Specifically, we focus on the impact of tertiary mass transfer on the evolution of the inner orbit and investigate whether it could lead to the formation of GW sources with distinct properties.}
   {To achieve this, we develop an analytical model that describes the evolution of the inner and outer orbits of hierarchical triples undergoing stable mass transfer from the tertiary component. We publicly release this model as a python package on Zenodo. Utilising population synthesis simulations, we investigate triples with a Roche-lobe filling tertiary star and an inner binary black hole (BBH). These systems stem from inner binaries experiencing chemically homogeneous evolution (CHE). Our analysis encompasses two distinct populations with metallicities of $Z=0.005$ and $Z=0.0005$, focusing on primary components in the inner binary with initial masses ranging from 20 to $100\,\text{M}_{\odot}$, and inner and outer orbital separations up to $40\,\text{R}_{\odot}$ and $10^5\,\text{R}_{\odot}$, respectively, targeting the parameter space where chemically homogeneous evolution is anticipated.}
 {Our results indicate that, for the systems studied, the mass transfer phase predominantly leads to orbital shrinkage of the inner binary and evolution towards non-zero eccentricities, accompanied by an expansion of the outer orbit. In the systems where the inner binary components evolve chemically homogeneous, 9.5\% result in mass transfer from the tertiary onto an inner BBH. Within this subset, we predict a high formation efficiency of GW mergers, ranging from 85.1\% to 100\% at $Z=0.005$ and 100\% at $Z=0.0005$, with short delay times, partly attributable to the mass transfer phase. Owing to the rarity of triples with a CHE inner binary in the stellar population, we project local merger rates in the range of 0.69 to 1.74 $\rm{Gpc^{-3} \ yr^{-1}}$. Of the prospected BBH mergers that enter  the LISA and aLIGO frequency band due to GW emission, a fraction is still accretion gas from the tertiary star. This could produce a strong electromagnetic (EM) counterpart to the GW source and maintain high eccentricities as the system enters the frequency range detectable by GW detectors. The occurrence of EM signals accompanying mergers varies significantly depending on model assumptions, with fractions ranging from less than 0.03\% to as high as 46.8\% of all mergers if the formation of a circumbinary disk is allowed.}
{}


   \maketitle
%

\section{Introduction}

Multiple stars are ubiquitous among the observed population of stars. For a detailed overview of stellar multiplicity see \citet{moe_mind_2017, offner_origin_2023}. Notably, approximately half of solar-type stars are found to possess at least one stellar companion \citep{tokovinin_binaries_2014-1}, with 10\% of all systems existing as triple systems \citep{tokovinin_binaries_2014}. As we move toward higher masses, the companion fraction increases. For young O-type stars, nearly every star is part of a multiple system, with almost 70\% comprising a triple or higher order multiple, although this estimate carries some uncertainty \citep{evans_vlt-flames_2005, evans_vlt-flames_2006, sana_binary_2012, sana_southern_2014, kobulnicky_toward_2014, moe_mind_2017, offner_origin_2023}. Given these statistics, the influence of stellar multiplicity on the evolution and ultimate fates of stars cannot be understated.

In hierarchical triples, unique dynamical, tidal and mass transfer interactions emerge through the interplay between the stars, phenomena not encountered in single- or binary-star evolution. For example, angular momentum exchange between the inner binary's orbit and the tertiary star can induce variations in the inner orbit's eccentricity and the mutual inclination of the triple, a mechanism known as the Von Zeipel-Lidov-Kozai (ZLK) mechanism \citep{von_zeipel_sur_1910, lidov_evolution_1962, kozai_secular_1962, naoz_eccentric_2016}. Smaller outer-to-inner period ratios could lead to higher eccentricities and merger/collision rates compared to the secular approximation \citep{gri18, fra19, man22}. Additionally, tidal deformation of the tertiary star can extract energy from the inner binary \citep{fuller_tidally_2013, gao_empirical_2020}. Furthermore, mass transfer from a tertiary proceeds in a fundamentally different manner compared to mass transfer in binary systems \citep{de_vries_evolution_2014, comerford_estimating_2020, glanz_simulations_2021}. While conventionally, only a small fraction of stars in hierarchical triples undergo tertiary mass transfer (TMT) \citep{hamers_double_2019, toonen_evolution_2020, kummer_main_2023}, recent studies by \citet{dorozsmai_stellar_2024} have shown that TMT in massive triples occurs in approximately 50\% of systems if the stars of the inner binary evolve chemically homogeneous. A comprehensive understanding of the system's evolution during TMT is therefore necessary to understand the final fate of these complex systems. 

CHE, initially presented within the framework of rapidly rotating stars \citep{maeder_evidences_1987}, was later proposed to be applicable as well to closely orbiting binaries \citep{de_mink_rotational_2009,song2016massive}. This mechanism suggests that stellar deformation caused by rapid stellar rotation in tidally locked systems with short orbital periods induces enhanced internal mixing, leading to a homogeneous chemical distribution across the stars . By continually replenishing the core with hydrogen from the envelope during the main sequence (MS), these stars avoid developing a core-envelope boundary, thus preventing significant expansion during their evolution. Consequently, CHE binaries have been proposed as potential progenitors of long gamma-ray bursts \citep{yoon2006single,aguilera2018related}, (pulsational) pair-instability supernovae \citep{du_buisson_cosmic_2020}, GW mergers, as they can maintain compact orbits while avoiding mass transfer \citep{de_mink_chemically_2016, mandel_merging_2016, du_buisson_cosmic_2020, hastings2020internal, riley_chemically_2021, dorozsmai_stellar_2024}, and potentially contributed to the reionisation of the Universe \citep{sibony2022impact, ghodla2023evaluating}. Analysis by \citet{tokovinin_tertiary_2006} indicates that in solar-type systems, short-period binaries are more often accompanied by a tertiary star than wider binaries, highlighting the growing importance of triples in the context of CHE. Despite being a theoretical implication of rapid rotation, observational evidence lends support to the plausibility of CHE, as systems with observed properties consistent with this evolutionary channel have been identified \citep[e.g.,][]{almeida_discovery_2015, shenar_tarantula_2017, sharpe_investigating_2024}. Although, \citet{abdul2019clues} did not find clear evidence of efficient mixing due to rotation.

In this paper, we aim to investigate the impact of stable TMT on the properties of the inner binary, along with exploring the population of GW mergers resulting from CHE evolution in triples. To achieve this, we create a rapid analytical model that simulates the evolution of a triple system during stable mass transfer from a tertiary, applicable to population synthesis simulations of hierarchical triple stars. We then apply the model to two population synthesis simulations performed by \citet{dorozsmai_stellar_2024} at metallicities of $Z=0.005$ and $Z=0.0005$.

In Sect. \ref{sec:tmt_model}, we describe the setup of the TMT model, followed by a discussion on the formation channel of TMT around an inner BBH and an overview of the initial conditions for the population synthesis simulations in Section \ref{sec:triple_pop}. We then provide a qualitative description of the evolution of an example system during the TMT phase for varying physical assumptions in Section \ref{sec:test_models}. In Sect. \ref{sec:results_population}, we present a statistical analysis of the evolution of the simulated populations during the TMT phase and the merger rates/properties of merging black holes. In Sect. \ref{sec:discussion}, we discuss caveats and assumptions of our study, and finally, in Sect. \ref{sec:conclusion}, we give a summary of our findings.

\section{Modelling tertiary mass transfer}
\label{sec:tmt_model}

In this section, we present our analytical framework designed to simulate stable mass transfer originating from a tertiary donor. The analytical nature of our model makes it well suited for integration within population synthesis simulations of triple systems. To provide necessary context, we outline the hierarchical triple star configuration under consideration. Such a system comprises an inner binary, characterised by primary and secondary stars possessing initial zero-age main-sequence (ZAMS) masses ($m_1$, $m_2$) and angular velocities ($\Omega_1$, $\Omega_2$) assuming rigid-body rotation, and an outer binary consisting of the tertiary star with mass $m_3$ and angular velocity $\Omega_3$, and the centre of mass of the inner binary. We define the orbital elements of the inner binary as $\{a_{\rm{in}},e_{\rm{in}},g_{\rm{in}}\}$, where $a_{\rm{in}}$ represents the semi-major axis, $e_{\rm{in}}$ the eccentricity and $g_{\rm{in}}$ the argument of pericenter. We establish a parallel set for the outer orbit denoted as $\{a_{\rm{out}},e_{\rm{out}},g_{\rm{out}}\}$. Lastly, we introduce the mutual inclination between the planes of the inner and outer orbit as $i_{\rm{mut}}$.

We divide this section in three parts: first, we describe the evolution of the inner binary throughout the TMT phase. Second, we describe the evolution of the outer orbit. Third, we discuss variations to the analytical model. The default model described in this section should be regarded as the most simplistic model, omitting numerous physical processes that are likely to take place. With each model variation, we introduce an extra physical process to the model to assess its significance. 

\subsection{Evolution of the inner binary}
\label{sec:evol_inner}
\subsubsection{Will a disk form?}

The trajectory of the mass stream originating from a tertiary star and approaching the inner binary can undergo two distinct evolutionary paths depending upon the stellar and orbital characteristics of the hierarchical triple system. Should the trajectory of the mass stream fail to intersect with the orbit of the inner binary, it is assumed that it will travel around the inner binary, intersect itself, and form a circumbinary disk (CBD) encompassing the binary. Conversely, if the mass stream crosses with the binary orbit, the formation of a CBD is prevented. In this study, we refer to this as ballistic accretion (BA). To determine the minimum distance, $R_{\rm{min}}$, at which the mass stream can approach the accreting binary, we adopt the fitting formulas from \citet{lubow_gas_1975, ulrich_accreting_1976}:
\begin{equation}
    R_{\rm{min}} = 0.425a_{\rm{out}}(1-e_{\rm{out}})\Bigg[\frac{1}{q_{\rm{out}}}\Bigg(1+\frac{1}{q_{\rm{out}}}\Bigg)\Bigg],
    \label{eq:cbd_vs_ba}
\end{equation}
where $q_{\rm{out}}$ is the mass ratio of the outer binary, defined as $q_{\rm{out}} = m_3/(m_1+m_2)$. If the apocenter of the inner binary, $a_{\rm{apo}}=a_{\rm{in}}(1+e_{\rm{in}})$, exceeds $R_{\rm{min}}$, we assume ballistic accretion occurs, whereas a CBD forms when the apocenter is smaller.

\subsubsection{Ballistic accretion}

During ballistic accretion, we assume that the gas from the stream is redistributed as it intersects with the binary orbit, forming a gas cloud encapsulating the inner binary. This structure bears resemblance to a common envelope phase, as found by \citet{de_vries_evolution_2014}. Recent analytical studies of stable mass transfer in hierarchical triples have treated these configurations as a classic common envelope, parameterised utilising the $\alpha\lambda$ formalism in order to predict the evolution of the inner binary \citep{hamers_statistical_2022, gao_stellar_2023, dorozsmai_stellar_2024}. In contrast, our methodology involves determining the evolution of the inner binary through the calculation of drag forces imparted on the binary components by the surrounding gaseous medium. Gravitational attraction of the wake trailing the objects remove linear momentum and slow down the objects. The amplitude of the force on a perturber mass ($m_p$) resulting from the gravitational drag force (GDF) was calculated by \citet{ostriker_dynamical_1999}: 
\begin{equation}
    \vec{F}_{\rm{GDF}} = -\frac{4\pi G^2m_p^2\rho_g}{v_{\rm{rel}}^3}\vec{v}_{\rm{rel}} \mathcal{I}(\mathcal{M}),
\end{equation}
with $\rho_g$ the density of the gas surrounding the perturber, $v_{\rm{rel}}$ the relative velocity between the gas and the perturber, and $\mathcal{I}(\mathcal{M})$ a dimensionless function dependent on the Mach number $\mathcal{M} = v_{\rm{rel}}/c_s$, where $c_s$ is the speed of sound in the gas. For lack of a better estimate, we make the assumption that the gas is stationary with respect to the center of mass (COM) frame. Consequently, the relative velocity is equivalent to the Keplerian velocity of the perturber. 
The value of $\mathcal{I(\mathcal{M})}$ has been determined by \citet{ostriker_dynamical_1999} for an object that moves on a straight trajectory through a uniform gaseous medium, either at subsonic $\mathcal{M}<1$ or supersonic $\mathcal{M}\geq1$ velocities:
\begin{equation}
\label{eq:lin_pert}
    I(\mathcal{M}) = 
    \begin{cases}
    \frac{1}{2}\rm{ln}\Bigg(\frac{1+\mathcal{M}}{1-\mathcal{M}}\Bigg) - \mathcal{M} & \text{if } \mathcal{M} < 1 \\
    \rm{ln}\Bigg(1-\frac{1}{\mathcal{M}^2}\Bigg) + \rm{ln}\Bigg(\frac{r_{\rm{max}}}{r_{\rm{min}}}\Bigg) & \text{if } \mathcal{M} \geq 1
    \end{cases}
\end{equation}
where $r_{\rm{max}}$ and $r_{\rm{min}}$ represent the characteristic sizes encompassed by the gravitational wake. In the case of a binary system, $r_{\rm{max}}$ typically approaches the orbital diameter \citep{kim_dynamical_2007, antoni_evolution_2019}, thus we assign $r_{\rm{max}}$ a value of $2a_{\rm{in}}$. While the exact value of $r_{\rm{min}}$ remains uncertain, we adopt the approach by \citet{kim_dynamical_2007} and set $r_{\rm{min}} = a_{\rm{in}}/10$, which makes $\ln(r_{\rm max}/r_{\rm min})=\ln20 \approx 3$.

The description of the wake formulated in \citet{ostriker_dynamical_1999} is under the assumption of a single object moving along a linear path through a gaseous medium. However, our situation involves two objects orbiting each other. To address this, \citet{kim_dynamical_2007} derived an expression for the gravitational wake of a perturber moving along a circular trajectory. Unlike the symmetrical wake in the former case, the wake has both radial and azimuthal components, with the azimuthal component dominating the orbital decay. To capture this effect, \citeauthor{kim_dynamical_2007} provide a fitting function for the azimuthal component of the wake:
\begin{equation}
\label{eq:wake_07}
    \mathcal{I_{\phi}^{\rm{K07}}(M)} = 
    \begin{cases}
        0.7706\ \rm{ln}\Bigg(\frac{1+\mathcal{M}}{1.0004-0.9185\mathcal{M}}\Bigg) -1.4703\mathcal{M} \\
        \hspace*{2.3cm} \text{if } \mathcal{M} < 1.0  \\[0.5em]
        \begin{array}{l}
        \rm{ln}(330(a_{\rm{in}}/r_{\rm{min}})\times(\mathcal{M}-0.71)^{5.72}\mathcal{M}^{-9.58}) \\
        \hspace*{2.1cm} \text{if } 1.0 \leq \mathcal{M} < 4.4 
        \end{array} \\[0.5em]
        \rm{ln}\Bigg(\frac{a_{\rm{in}}/r_{\rm{min}}}{0.11\mathcal{M}+1.65}\Bigg) \quad \text{if } 4.4 \leq \mathcal{M} 
    \end{cases}
\end{equation}
In the case of a binary system, the wake generated by one component influences the gravitational drag experienced by the other component. To address this mutual interaction, \citet{kim_dynamical_2008} introduces an additional term to the gravitational wake expression:
\begin{equation}
\label{eq:wake_08}
    \frac{\mathcal{I_{\phi}^{\rm{K08}}(M)}}{\mathcal{M}^2} = 
    \begin{cases}
        -0.022(10-\mathcal{M}) \tanh(3\mathcal{M}/2) & \text{if } \mathcal{M} < 2.97  \\
        -0.13+0.07\rm{tan^{-1}}(5\mathcal{M}-15) & \text{if } \mathcal{M} \geq 2.97  \\
    \end{cases}
\end{equation}
To obtain the total effect of the wake, we sum the contributions from both terms, $\mathcal{I}_{\phi}(\mathcal{M}) = \mathcal{I_{\phi}^{\rm{K07}}(M)}+\mathcal{I_{\phi}^{\rm{K08}}(M)}$.

In Appendix \ref{sec:appA} we derive the change in semi-major axis and eccentricity of the inner orbit over time. For a circular orbit, the eccentricity remains constant, and the semi-major axis decreases due to the dissipation of orbital energy \citep{grishin_application_2016,rozner_binary_2022}:
\begin{equation}
\label{eq:da_gdf}
    \frac{da_{\rm{in}}}{dt}\Big|_{\rm{BA}} = -8\pi\sqrt{\frac{Ga^5}{M}}\rho_g\mathcal{I}\Bigg[\frac{1}{q}(1+q^{-1})^2 + q(1+q)^2\Bigg],
\end{equation}
where $M=m_1+m_2$ is the total mass of the inner binary and $q$ is the mass ratio of the inner binary, defined as $q=m_2/m_1$.

To simulate the evolution of the inner binary, we have to make assumptions regarding the gas density and local sound speed around the perturbers. However, detailed simulations providing realistic values for these parameters in such systems are currently lacking. Especially, the gas density is anticipated to have a considerable influence on the rate of evolution of the inner binary, given that the drag force is directly proportional to $\rho_g$. To address this uncertainty, we evaluate $F_{\rm{GDF}}$ at two distinct gas density values: $\rho_g=10^{-8}\rm{g \ cm^{-3}}$ and $\rho_g=10^{-10}\rm{g \ cm^{-3}}$, ensuring that these densities are less dense compared to typical common envelope scenarios in binaries or triples, which are about $10^{-5} \ \rm{g \ cm^{-3}} - 10^{-7} \ \rm{g \ cm^{-3}}$ \citep[e.g.,][]{ivanova_common_2013, glanz_simulations_2021}. For simplicity, we assume a constant gas density over time, implying that the mass transfer rate onto the inner binary equals the amount of mass expelled from the binary. Unless otherwise specified, we assume that none of the mass is accreted by the binary components. Additionally, we adopt a sound speed value of $\rm{30 \ km\ s^{-1}}$, corresponding to a gas temperature of approximately $\rm{10^5 \ K}$.

\subsubsection{Circumbinary disks}

The dynamics governing the interactions between a binary system and the surrounding CBD are inherently intricate. For a detailed review on CBDs and their applications, we direct the readers to the comprehensive review by \citet{lai_circumbinary_2022}. Various processes can significantly influence the orbital evolution of the binary. On one hand, gravitational torques stemming from resonant interactions between the binary and the disk can extract angular momentum from the binary. Conversely, accretion from the disk onto the binary may transfer angular momentum inward. The balance of these torques dictate how the inner orbit evolves over time. The configuration of the binary, characterised by parameters such as mass ratio ($q_{\rm{in}}$), eccentricity ($e_{\rm{in}}$), and mutual inclination ($i_{\rm{mut}}$), appears to play a pivotal role in determining the binary's fate \citep{nixon_tearing_2013, miranda_tidal_2015, miranda_viscous_2017, munoz_hydrodynamics_2019, munoz_circumbinary_2020, duffell_circumbinary_2020, zrake_equilibrium_2021,  dorazio_orbital_2021, siwek_preferential_2023, siwek_orbital_2023}. Furthermore, alterations in binary properties can lead to consequential shifts in its evolutionary trajectory \citep{valli_long-term_2024}.

\begin{figure}
    \centering
    \begin{subfigure}{\linewidth}
        \includegraphics[width=\linewidth]{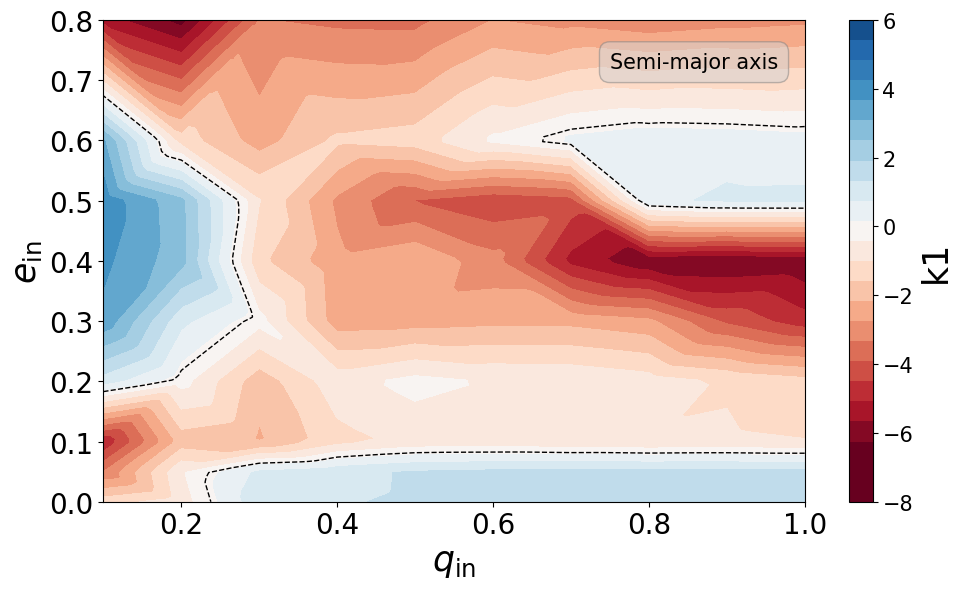}
        \label{fig:adot_cbd}
    \end{subfigure}
    \hfill
    \begin{subfigure}{\linewidth}
        \includegraphics[width=\linewidth]{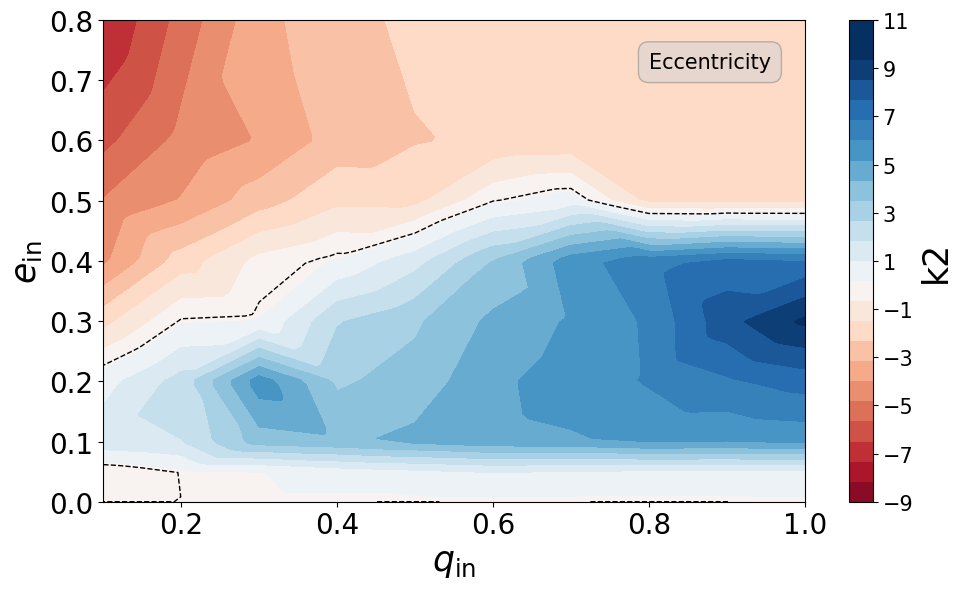}
        \label{fig:edot_cbd}
    \end{subfigure}
    \caption{Constants for the time derivatives of the semi-major axis (top) and the eccentricity (bottom) of an inner binary surrounded by a CBD, based on the findings of \citet{siwek_orbital_2023}. The figure illustrates how certain parts of the parameters space lead to either a decrease or increase of the semi-major axis and the eccentricity. The dashed lines indicate the points the derivative equals zero.}
    \label{fig:cbd_deriv}
\end{figure}

We utilise findings from \citet{siwek_orbital_2023}, who performed hydrodynamical simulations of CBDs around super-massive black holes (SMBHs). Regarding the disk properties, a constant value of 0.1 is assigned to the disk viscosity parameter $\alpha$. Additionally, the aspect ratio $h$, representing the disk's thickness relative to its size, is fixed at 0.1. The study encompasses an extensive parameter exploration using a grid-based approach, wherein the eccentricity, $e_{\rm{in}}$, ranges from 0 to 0.8, and the mass ratio, $q_{\rm{in}}$, ranges from 0.1 to 1. For each grid point, time derivatives for the semi-major axis and eccentricity are provided. The rate of change of $a_{\rm{in}}$ and $e_{\rm{in}}$ is expressed in terms of the mass accretion rate and binary mass, as follows:
\begin{equation}
\label{eq:da_cbd}
    \frac{da_{\rm{in}}}{dt}\Big|_{\rm{CBD}} = k_1\frac{\dot{m}_{\rm{bin}}}{m_{\rm{bin}}}a_{\rm{in}},
\end{equation}
\begin{equation}
\label{eq:de_cbd}
    \frac{de_{\rm{in}}}{dt}\Big|_{\rm{CBD}} = k_2\frac{\dot{m}_{\rm{bin}}}{m_{\rm{bin}}}e_{\rm{in}},
\end{equation}
where $k_1$ and $k_2$ are factors dependent on $q_{\rm{in}}$ and $e_{\rm{in}}$, and $\dot{m}_{\rm{bin}}$ is always positive as the binary is accreting material. We use linear interpolation to determine the values of $k_1$ and $k_2$ for any given $q_{\rm{in}}$ and $e_{\rm{in}}$, as illustrated in Fig. \ref{fig:cbd_deriv}. A noteworthy observation is the existence of an equilibrium eccentricity around $e_{\rm{in}}=0.45-0.5$ for near-equal mass ratios, likely attributed to a resonant locking mechanism \citep{artymowicz_effect_1991}. For $e_{\rm{in}}>0.8$, extrapolation is avoided, and instead, the values for $e_{\rm{in}}=0.8$ are adopted.

In scenarios where the orbital configuration permits the formation of a CBD, we assume that the disk promptly reaches a steady state. Here, the amount of mass accreted onto the inner binary is continuously replenished by the mass transferred from the tertiary star. We adopt the fitting formula proposed by \citet{siwek_preferential_2023} for the preferential accretion onto each binary component, which aligns well with the outcomes of other hydrodynamic simulations of CBDs surrounding SMBHs \citep{munoz_circumbinary_2020, duffell_circumbinary_2020}. This relationship is expressed as:
\begin{equation}
    \frac{\dot{m}_2}{\dot{m}_1} = \Bigg(\frac{m_2}{m_1}\Bigg)^{-0.9},
\end{equation}
resulting in a tendency towards mass equalisation within the binary system.

\subsubsection{Gravitational waves}

At small separations, angular momentum loss through emission of gravitational waves becomes non-negligible for binary compact objects and can assist the inspiral of the binary towards a merger. The evolution of the semi-major axis and the eccentricity are described by \citet{peters_gravitational_1964}:
\begin{align}
    \frac{da_{\rm{in}}}{dt} = & \ \frac{64}{5}\frac{G^3m_1m_2(m_1+m_2)}{c^5a_{\rm{in}}^3(1-e_{\rm{in}}^2)^{7/2}}\Bigg[1+e_{\rm{in}}^2\frac{73}{24}+e_{\rm{in}}^4\frac{37}{96}\Bigg] \nonumber \\[10pt]
    \frac{de_{\rm{in}}}{dt} = & \ \frac{304}{15}\frac{G^3m_1m_2(m_1+m_2)}{c^5a_{\rm{in}}^4(1-e_{\rm{in}}^2)^{5/2}}\Bigg[1+e_{\rm{in}}^2\frac{121}{304}\Bigg]
\end{align}

The complete evolution of $a_{\rm{in}}$ and $e_{\rm{in}}$ will be governed by the summed contribution of the gas drag during ballistic accretion, CBD dynamics, and gravitational wave emission
\begin{align}
    \frac{da_{\rm{in}}}{dt}\Big|_{\rm{total}} = & \ \frac{da_{\rm{in}}}{dt}\Big|_{\rm{BA}} + \frac{da_{\rm{in}}}{dt}\Big|_{\rm{CBD}} + \frac{da_{\rm{in}}}{dt}\Big|_{\rm{GW}} \nonumber \\[10pt]
    \frac{de_{\rm{in}}}{dt}\Big|_{\rm{total}} = & \ \frac{de_{\rm{in}}}{dt}\Big|_{\rm{CBD}} + \frac{de_{\rm{in}}}{dt}\Big|_{\rm{GW}}
\end{align}
When the TMT phase proceeds via ballistic accretion, the contribution of the CBD is zero, and vice versa.  

\subsubsection{Eccentric gas drag}

Initially, we considered gas drag effects on objects moving along circular orbits, leading to a constant time derivative of the semi-major axis along the orbit and a constant eccentricity of zero. However, three-body dynamical effects, such as ZLK cycles can excite the eccentricity of the inner binary prior to the onset of mass transfer. The drag forces become phase dependent, as the velocity of the binary objects relative to the gas is different between periastron and apastron. In Appendix \ref{sec:appA}, we derive the orbit-averaged time derivatives of the semi-major axis and eccentricity. For binaries embedded in a homogeneous gas cloud, the GDF leads to an increase in eccentricity \citep[e.g.,][]{szolgyen_eccentricity_2022}. \citet{rozner_binary_2022} demonstrate that a steep growth of eccentricity towards unity can be achieved, potentially facilitating a merger of the binary. 

\subsubsection{Isotropic re-emission}

So far we have assumed that the components of the inner binary are able to accrete all material that is transferred from the CBD. However, in reality, the accretion rate is likely to be limited. For stars, accretion efficiency is often tied to the star's thermal timescale, while for compact objects, it is constrained by their Eddington limit. If the mass transfer rate exceeds the maximum accretion rate, any excess material is expelled from the vicinity of the accreting object. Therefore, we explore a scenario where none of the material is accreted and is instead ejected from the system in the form of a fast isotropic wind or accretion disk wind \citep[e.g.,][]{begelman_compton_1983,gallegos2023angular}. The loss of angular momentum due to this expulsion affects the orbital evolution of both the inner and outer orbits in addition to the evolution resulting from CBD interactions. We assume the matter is expelled symmetrically with respect to the object and the evolution of the inner binary is similar to that of an isotropic outflow \citep{hadjidemetriou_two-body_1963, dosopoulou_orbital_2016}: 
\begin{align}
\label{eq:da_de_iso}
    \frac{da_{\rm{in}}}{dt} = & \ -a_{\rm{in}}\frac{\dot{m}^{\text{iso}}_1+\dot{m}^{\text{iso}}_2}{m_1+m_2} \nonumber \\[10pt]
    \frac{de_{\rm{in}}}{dt} = & \ 0,
\end{align}
where $\dot{m}^{\text{iso}}$ is the mass lost via an isotropic outflow. It is important to note that $a_{\rm{in}}$ increases over time due to the inner binary losing mass, with $\dot{m_1},\dot{m_2}<0$. The consequences of re-emission on the outer orbit will be discussed later.

\subsubsection{Retrograde torques}

Thus far, our model for the CBD interactions assumed that the CBD rotates prograde with respect to the inner binary. However, triples are observed to have mutually inclined inner and outer orbits \citep{borkovits_comprehensive_2016}. If the mutual inclination is larger than $90^\circ$, the tertiary star will orbit in retrograde motion with respect to the inner binary, which may result in the formation of a retrograde disk. \citet{nixon_retrograde_2011} demonstrated that for retrograde disks, orbital resonances are absent. Additionally, the binary components accrete negative angular momentum as the angular momentum vector of the gas particles in a retrograde disk points the opposite direction, leading to a rapid inspiral of the inner binary. \citet{tiede_eccentric_2023} performed an extensive parameter study for retrograde disks around SMBH binaries and provided an analytical fit based on their numerical results for the evolution of the semi-major axis and eccentricity:
\begin{align}
\label{eq:da_de_retro}
    \frac{da_{\text{in}}}{dt} = & \ -10a_{\text{in}}\frac{\dot{m}_{\text{bin}}}{m_{\text{bin}}} \nonumber \\[10pt]
    \frac{de_{\text{in}}}{dt} = & \ 30 \max\{0.1,e_{\text{in}}\} \frac{\dot{m}_{\text{bin}}}{m_{\text{bin}}}
\end{align}

\subsection{Evolution outer binary}

The orbital evolution of the outer binary is governed by the exchange and loss of angular momentum within the triple system. Assuming that the tertiary star is sufficiently distant from the inner binary, we can simplify the system by treating the inner binary as a point mass. Under this approximation, the evolution of the semi-major axis of the outer orbit is described by the equation \citep{soberman_stability_1997, hamers_multiple_2021}:
\begin{equation}
\label{eq:da_out}
    \frac{da_{\rm{out}}}{dt} = -2a_{\rm{out}}\frac{\dot{m}_3}{m_3}\Bigg[1-\beta\frac{m_3}{m_1+m_2}-(1-\beta)(\gamma+\frac{1}{2})\frac{m_3}{m_1+m_2+m_3}\Bigg],
\end{equation}
analogous to mass transfer in isolated binaries. Here, $\beta$ is the mass accretion efficiency onto the accretor (i.e., the inner binary), where $\beta=0$ and $\beta=1$ represent completely non-conservative and conservative mass transfer, respectively. The parameter $\gamma$ represents the specific angular momentum of the material that is expelled from the system. For simplicity, we assume that the eccentricity remains unchanged during the mass transfer.

In our fiducial model, we assume that all transferred material during ballistic accretion is eventually expelled isotropically. This material carries a specific angular momentum equal to that of the inner binary. Thus, we set $\beta=0$ and $\gamma=m_3/(m_1+m_2)$. If a CBD forms, we assume that all transferred mass is accreted onto the inner binary. In this scenario, $\beta=1$ (and $\gamma$ remains unspecified). 

In scenarios involving isotropic re-emission within the CBD, the mass transfer is considered entirely non-conservative, such that $\beta=0$ and $\gamma=m_3/(m_1+m_2)$. Concerning the evolution of the outer orbit, we assume that all material is ejected at the centre of mass of the inner binary, and therefore carries the specific angular momentum of the inner binary. This assumption is justified by the substantial difference in size between the outer and inner orbits.

\subsection{Model variations}
\label{sec:variations}

Here we describe the various models that we will investigate later in the paper. A summary of all models is presented in Table \ref{tab:model_assumptions}. First, we start with the most simplistic scenario, only considering BA and GW emission. This will be called Model Simple. The GDF exerted on the binary components is based on the linear perturbation theory (Eqs. \ref{eq:lin_pert}+\ref{eq:da_gdf}). Estimating these forces requires making assumptions for the gas density and local sound speed around the binary components. To evaluate the implications of the uncertainty in the gas properties, we run 4 models with varying gas densities and mass transfer rates (defined in Sect. \ref{sec:stability}), comparing the resulting final inner orbits across the models. Additionally, we include the mass-transfer rate as a parameter, as it is expected to influence the gas density around the inner binary. Thus, we explore all four permutations of two gas densities, $\rm{\rho_{gas}=10^{-8}g\ cm^{-3}}$ and $\rm{\rho_{gas}=10^{-10}g\ cm^{-3}}$, and two mass-transfer rates, namely the mass transfer that occurs on the thermal and the nuclear timescale of the donor star. 

For the next model, we introduce the potential formation of a CBD around the inner binary. The evolution of the semi-major axis and eccentricity in the presence of a CBD are described by Eqs. \ref{eq:da_cbd} \& \ref{eq:de_cbd}. We assume the mass transfer onto the components of the inner binary is conservative. We refer to this as Model Basic. Similar to the previous model, we explore the same four combinations of gas densities and mass transfer rates. However, in Section \ref{sec:evol_a}, we demonstrate that in the presence of a CBD, the final distribution of semi-major axes in a population tends to converge with increasing gas density. Therefore, for subsequent models, we focus solely on the combination of $\rm{\rho_{gas}=10^{-8}g\ cm^{-3}}$ with nuclear timescale mass transfer, and $\rm{\rho_{gas}=10^{-10}g\ cm^{-3}}$ with thermal timescale mass transfer, representing the two extremes.

The following four models extend Model Basic by incorporating additional physical phenomena, with each model focusing on one aspect at a time to assess its significance. Model Basic+BinGDF integrates the non-symmetric gravitational drag forces experienced by binary components during BA, as described by Eqs. \ref{eq:wake_07} \& \ref{eq:wake_08}, originating from the wake generated by two binary components. Model Basic+EccGDF introduces a modification to the gravitational drag forces under the assumption of eccentric orbits during BA. The evolution of semi-major axis and eccentricity follows Eqs. \ref{eq:app_da} \& \ref{eq:app_de}. In Model Basic+Iso, we assume that during the CBD phase, none of the transferred mass accretes onto the inner binary components, instead being isotropically (or symmetrically) ejected near the accreting object. An additional component of the orbital evolution due to the angular momentum loss from the inner binary is described by Eq. \ref{eq:da_de_iso}. Model Basic+Retro incorporates an adjusted formalism for the evolution of semi-major axis and eccentricity during the CBD phase for systems with a retrograde disk, described by Eq. \ref{eq:da_de_retro}. Finally, we merge the Basic+ models, incorporating all four additional physical processes described in Section \ref{sec:evol_inner}. We refer to this as Model Advanced. We emphasise that Model Advanced is not necessarily the most realistic model, as the physical assumptions are subject to uncertainties.

We introduce additional variations to the Basic and Advanced models, referred to as Basic+NoDrag and Advanced+NoDrag, where the effects of gas drag are completely ignored. If the total mass of the gaseous cloud enveloping the inner binary is significantly lower than that of the binary components, the formation of an overdense wake trailing the objects becomes inefficient, leading to a minimal inspiral of the binary. In this scenario, we specifically examine the case where the inspiral process is entirely ineffective.

\begin{table*}
\caption{Physical assumptions for the different models explored in this study. Combinations of gas densities and mass transfer rates: A = [$\rho_g=10^{-8} \ \& \ \dot{m}_3=\dot{m}_{\rm{nuc}}$], B = [$\rho_g=10^{-8} \ \& \ \dot{m}_3=\dot{m}_{\rm{th}}$], C = [$\rho_g=10^{-10} \ \& \ \dot{m}_3=\dot{m}_{\rm{nuc}}$], D = [$\rho_g=10^{-10} \ \& \ \dot{m}_3=\dot{m}_{\rm{th}}$].}
    \centering
    \begin{tabular}{|l|m{0.7cm}|p{0.7cm}|p{0.7cm}|p{0.7cm}|p{0.7cm}|m{0.7cm}|l|}
        \hline
        \diagbox[width=10.5em,height=3.35em]{\textbf{Model}}{\textbf{Physics}} & \centering BA & \centering Bin. GDF & \centering Ecc. GDF & \centering CBD & \centering Iso. & \centering Retro. & $\rho_g \ \& \ \dot{m}_3$  \\
        \hline\hline
        Model Simple & \centering x & & & & & & A, B, C, D \\ \hline
        Model Basic & \centering x & & & \centering x & & & A, B, C, D \\ \hline
        Model Basic+BinGDF & \centering x & \centering x & & \centering x & & & A, D \\ \hline
        Model Basic+EccGDF & \centering x & & \centering x & \centering x & & & A, D \\ \hline
        Model Basic+Iso & \centering x & & & \centering x & \centering x & & A, D \\ \hline
        Model Basic+Retro & \centering x & & & \centering x & & \centering x & A, D \\ \hline
        Model Advanced & \centering x & \centering x & \centering x & \centering x & \centering x & \centering x & A, D \\
        \hline
        Model Basic+NoDrag & & & & \centering x & & & A, D \\
        \hline
        Model Advanced+NoDrag & & & & \centering x & \centering x & \centering x & A, D \\
        \hline
    \end{tabular}
    \label{tab:model_assumptions}
\end{table*}

\section{Triple population synthesis}
\label{sec:triple_pop}

\subsection{Preceding secular evolution}
\label{sec:sec_evol}

\begin{figure}
    \includegraphics[width=\hsize]{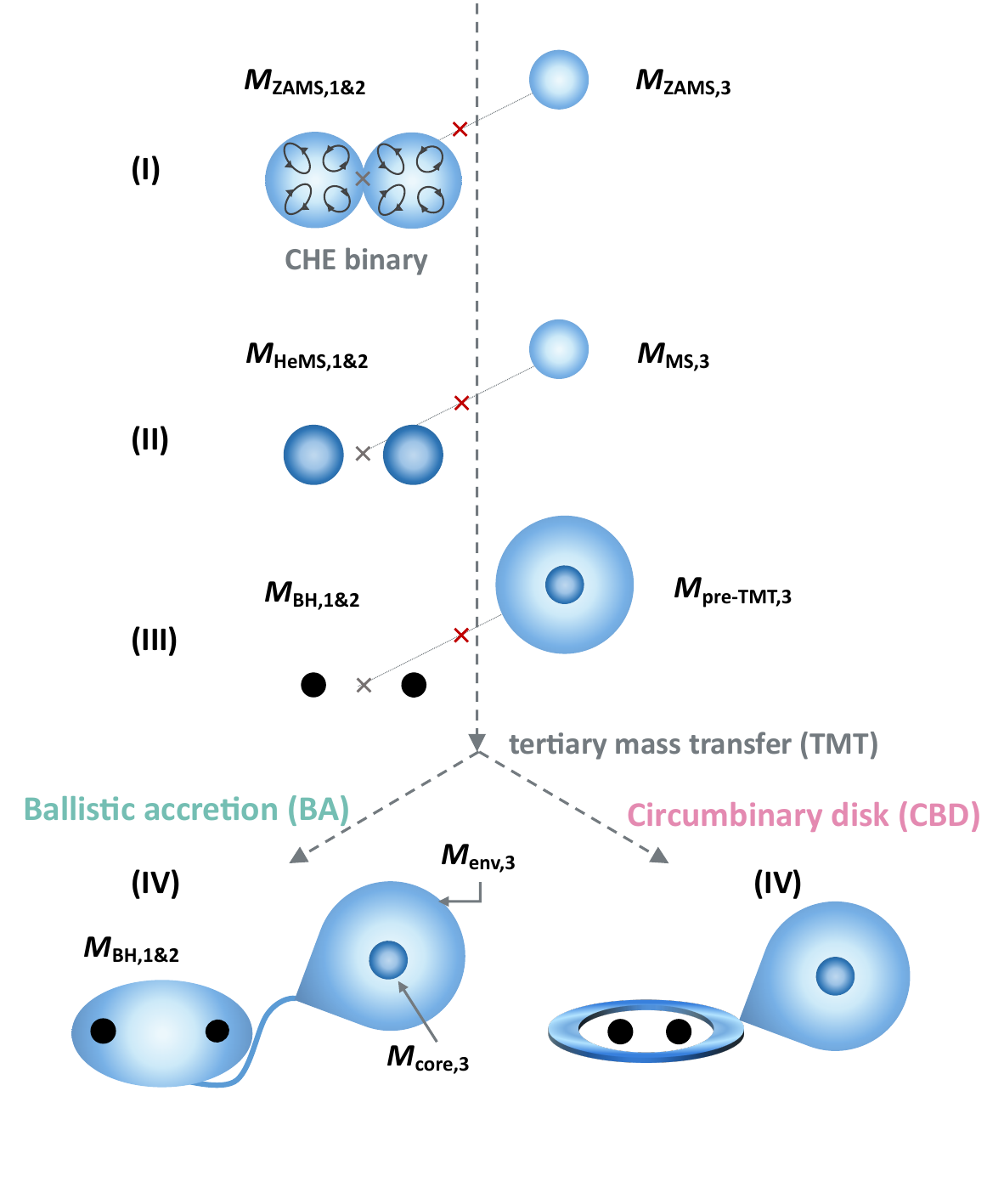}
    \caption{A cartoon illustrating the evolutionary stages of triples with a CHE inner binary that result in TMT with an inner BBH. Depending on the properties of the triple at the onset of TMT, mass transfer occurs via BA or the formation of a CBD. The masses of the three companions are indicated at the zero-age main sequence ($M_{\rm{ZAMS}}$), main sequence ($M_{\rm{MS}}$) and helium main sequence ($M_{\rm{HeMS}}$) phases. Furthermore, the black-hole mass ($M_{\rm{BH}}$), pre-tertiary mass-transfer mass ($M_{\rm{pre-TMT}}$), envelope mass ($M_{\rm{env}}$) and core mass ($M_{\rm{core}}$) are indicated. The grey and red crosses denote the centre of mass of the inner binary and outer binary, respectively. The figure is not to scale and has been adapted from \citet{van_son_redshift_2022}.}
    \label{fig:cartoon_tmt}
\end{figure}

We first give an overview of the typical evolution of a triple system with a CHE inner binary that leads to TMT and highlight the important intermediate phases. For the system to potentially produce a GW merger, we implement the condition that the components of the inner binary have already evolved to their remnant stage prior to the onset of mass transfer. A cartoon of the evolution is provided in Figure \ref{fig:cartoon_tmt}. 

(I) At the ZAMS, the stars of the inner binary are on a sufficiently short orbit, such that tidal dissipation is strong enough to have synchronised the rotational period of the stars with the orbital period, commonly referred to as tidal locking. Consequently, rotation-induced mixing can prevent the star from establishing a chemical gradient \citep{maeder_evidences_1987, maeder_evolution_2000, heger_presupernova_2000}. As opposed to classical evolving stars, these CHE stars do not increase their radius during the MS. On the contrary, by preventing the formation of a chemical gradient within the star and enriching the envelope with helium, the star is expected to contract \citep{yoon_single_2006, mandel_merging_2016}. Due to the short orbital period, the binary is likely to be in over-contact. During this phase, either star must prevent from overflowing its second Lagrangian (L2) point, as this would lead to loss of angular momentum through mass outflow, followed by a merger of the binary \citep{soberman_stability_1997,marchant_new_2016}. During the MS phase of the inner binary components, any three-body dynamical effects are effectively quenched by strong tidal forces, preventing, e.g., an eccentric merger from occurring. Initially, the tertiary is less massive than either component of the inner binary, ensuring the star will not significantly expand and fill its Roche lobe before the other stars have formed a BH.

(II) Both stars of the inner binary have completed hydrogen fusion. As the interior of the stars has been efficiently mixed during the MS phase, the entire hydrogen content has been fused into helium. The radius of such a helium star is smaller compared to the radius at the ZAMS, and hence the stars detach and stop evolving chemically homogeneous, with the continued evolution similar to that of a helium star. As tidal forces are not as strong as before, dynamical interactions between the tertiary and the inner binary can become more pronounced. 

(III) The stars of the inner binary undergo core collapse and a BBH is formed. The supernovae might be accompanied by a kick, mainly leading to an increase in eccentricity of both the inner and the outer orbit, but not strong enough to unbind the system. Eventually, the tertiary will evolve off the MS and expand until it fills its Roche lobe, typically when evolving through the Hertzsprung Gap, or after helium has been ignited in the core. We stress that the radial evolution of the most massive stars remains poorly understood, and constitutes a major uncertainty in determining whether the tertiary star can fill its Roche lobe.

(IV) The tertiary star commences mass transfer toward the inner binary. The TMT can proceed in two distinct ways: (i) if the inner orbit is wide and/or the outer orbit is compact, the mass stream intersects with the orbit resulting in ballistic accretion. (ii) if the inner orbit is compact and/or the outer orbit is wide, the mass stream misses the orbit and intersects with itself, forming a CBD.

\subsection{Initial model assumptions}

We apply the TMT model described in Section \ref{sec:tmt_model} on a large number of hierarchical triple systems with a Roche-lobe filling tertiary star. The prior evolution of each system is simulated with the triple population synthesis code TRES\footnote{This code is publicly available on: \url{https://github.com/amusecode/TRES}} \citep{toonen_evolution_2016}. In total, we evolve $6\times10^5$ systems from the ZAMS, spread over metallicities of $Z=0.005$ (0.25 Z$_{\odot}$) and $Z=0.0005$ (0.025 Z$_{\odot}$), identical to the set-up described in \citet{dorozsmai_stellar_2024}. We only select systems that follow the evolutionary steps described in the previous section. In the remainder of this section, we provide a description of the initial set-up of the simulations and the conditions that ensure CHE in the inner binary.

The ZAMS properties of the stars and orbits are based on surveys of massive binaries \citep{sana_binary_2012, kobulnicky_toward_2014} for the inner binaries and observations of solar-type stars in hierarchical triples \citep{tokovinin_tertiary_2006, tokovinin_binaries_2014} for the outer binaries. The initial masses of the primary star, $m_1$, follow a power-law distribution with index $\alpha = -2.3$ \citep{kroupa_variation_2001}. The initial mass ratios between of the inner orbit, $q_{\rm{in}}=m_2/m_1$, and the outer orbit, $q_{\rm{out}}=m_3/(m_1+m_2)$, both follow a flat distribution. The initial inner and outer semi-major axes follow a flat distribution in log-space. The inner orbits are initially circular and the stellar spins are synchronised with the orbital period, consistent with being tidally locked. The outer eccentricities are drawn from a thermal distribution. The arguments of pericentre of the inner and outer binary are uniformly sampled between $-\pi$ and $\pi$. Lastly, the relative inclination is sampled uniformly from a cosine distribution between 0 and $\pi$. 

To ensure that the stars of the inner binary are tidally locked and can rotate rapidly enough to sustain efficient mixing within the star, we sample the initial system properties from a restricted range of values. The initial mass of the primary stars are within the range $\rm{[20,100]\ M_{\odot}}$ and the semi-major axis of the inner orbits are within the range $\rm{[14,40]\ R_{\odot}}$. Additionally, following \citet{dorozsmai_stellar_2024} to prevent an early stellar merger, the mass ratios of the inner binary are sampled between 0.7 and 1. The ranges for the outer binary are more relaxed: the tertiary masses are in the range $\rm{[5,100]\ M_{\odot}}$, mass ratios in the range [0.1,1] and the semi-major axes in the range $\rm{[100,10^5]\ R_{\odot}}$. 

We determine for each sampled system whether CHE takes place in the stars of the inner binary by comparing their rotational angular velocity to a critical
angular velocity that is necessary for efficient internal mixing. We use fits produced by \citet{riley_chemically_2021} that are based on numerical simulations performed by \citet{marchant_new_2016}, that provide these critical angular velocities for the given set of parameters of the inner binary. Any binary whose stellar components do not have an angular velocity exceeding this critical limit at the ZAMS, or at any later stage of the MS evolution, are discarded. During the CHE, the radius remains constant, apart from the effects of mass loss due to stellar winds. Once a CHE star leaves the MS, it is assumed to directly transition to a helium star. Following the Hurley tracks for MS and helium stars \citep{hurley_comprehensive_2000}, this coincides with a sudden drop in radius of the star. However, detailed simulations of chemically homogeneously evolving systems predict a more gradual decrease of the radius over the MS \citep[see][]{maeder_evidences_1987}. 

As the initial separation of the inner binary is small, the binary is likely to be in contact. During the contact phase, we assume that the masses equalise instantaneously \citep[see][]{riley_chemically_2021}; although observations suggest that contact binaries are preferably in non-equal mass configurations \citep{menon_detailed_2021, abdul-masih_constraining_2022}, which could be an indication that theoretical models of contact binaries are missing some important physics \citep{abdul2021constraining,fabry2023modeling}. Due to the redistribution of angular momentum within the binary, the semi-major axis evolves as:
\begin{equation}
   \frac{a_f}{a_i} = \Bigg(\frac{m_{1,i}m_{2,i}}{m_{1,f}m_{2,f}} \Bigg)^2,
\end{equation}
where the subscripts $i$ and $f$ refer to the initial and final state of the mass equalisation. If either star overflows their L2 point, the systems is discarded. The location of the L2 point is modelled following \citet{marchant_new_2016}:
\begin{equation}
\label{l2}
    \frac{R_{L2,2}-R_{RL,2}}{R_{RL,2}} = 0.299\tan^{-1}(1.84q^{0.397}),
\end{equation}
where $R_{RL,2}$ is the Roche lobe of the secondary star. \citet{}

\subsection{Stability of mass transfer}
\label{sec:stability}
The stability of the TMT is an important ingredient for the outcome of the mass-transfer phase. CE evolution in triples has been associated with dynamical ejections through destabilisation of the triple \citep[e.g.,][]{comerford_estimating_2020}; the plunge-in timescale of the inner binary towards the core of the tertiary star is short compared to the orbital evolution of the inner binary. Although, other outcomes such as a merger of the inner binary or a merger between the evolved tertiary star and the inner binary are also possible \citep{glanz_simulations_2021}. To that end, we only consider systems where the TMT phase is dynamically stable. \citet{dorozsmai_stellar_2024} predicts the vast majority of TMT phases to be dynamically stable as the mass ratios between the tertiary star and the inner binary ($q_{\rm{out}}$) are typically smaller than 1. 

The stability of the mass transfer phase is determined by the evolution of the Roche lobe and the radius of the donor star as response to mass loss. As the star tries to regain hydrostatic and thermal equilibrium, the radius might shrink or expand, depending on the structure of the envelope \citep{hjellming_thresholds_1987, soberman_stability_1997}. We determine the logarithmic derivatives of the radius with respect to the mass on the dynamical timescale:
\begin{equation}
\zeta_{\text{ad}} = \left( \frac{\text{d} \ln R}{\text{d} \ln m} \right)_{\text{ad}}
\label{eq_zeta_ad}
\end{equation}
The value of $\rm{\zeta_{ad}}$ depends on the structure of the envelope, and thus on the evolutionary type of the donor star. The values used for each evolutionary type is similar to those described in the binary population synthesis code SeBa \citep{toonen_supernova_2012}. The evolution of the Roche lobe during the mass transfer phase is mainly governed by the redistribution of mass within the binary and eventual angular momentum loss. We calculate $\rm{\zeta_{RL}}$:
\begin{equation}
\zeta_{\text{RL}} = \left( \frac{\text{d} \, \ln R_{\text{L}}}{\text{d} \, \ln m} \right).
\label{eq_zeta_rl}
\end{equation}
The mass transfer is dynamically stable when $\rm{\zeta_{RL}} \leq \zeta_{ad}$, i.e., the star can restore hydrostatic equilibrium. 

For dynamically stable mass transfer, the rate of mass loss from the donor star depends on the thermal response of its envelope. If the star is able to regain thermal equilibrium, the mass transfer occurs on the nuclear timescale of the donor star. The corresponding mass transfer rate can be estimated as follows:
\begin{equation}
\label{eq:m_nuc}
    \dot{m}_{\rm{nuc}} = \frac{m}{\tau_{\rm{nuc}}} \approx \frac{10^{-10}}{\phi}\frac{L}{\rm{L}_{\odot}}\rm{M_{\odot}yr^{-1}},
\end{equation}
where $\phi$ is a factor depending on the type of nuclear burning. For core-hydrogen burning, $\phi=1$. For core-helium burning, we assume $\phi=0.1$. If the star does not regain thermal equilibrium, the mass transfer occurs on its thermal timescale. The corresponding mass transfer rate can be estimated as follows:
\begin{equation}
\label{eq:m_th}
 \dot{m}_{\rm{th}} = \frac{m}{\tau_{\rm{th}}} \approx \frac{2RL}{\text{G} m} \approx 6.67\times10^{-8}\left(\frac{\rm{M}_{\odot}}{m}\right)\left(\frac{R}{\rm{R}_{\odot}}\right)\left(\frac{L}{\rm{L}_{\odot}}\right)\rm{M_{\odot}yr^{-1}},
\end{equation}
with $\tau_{\rm{th}}$ the thermal timescale and $L$ the luminosity of the donor star.

In our simulations, if the mass transfer is dynamically stable, we either assume mass transfer on the nuclear timescale or the thermal timescale for all systems. This way, we can more effectively capture the full range of uncertainties of the mass transfer on the formation of GW mergers.

\section{Example TMT model}
\label{sec:test_models}

We apply our model to a representative system drawn from the population outlined in Sect. \ref{sec:triple_pop}, mainly focusing on the evolution of the inner binary. To highlight the effects from the different TMT models, we qualitatively compare the differences between all the models described in Sect. \ref{sec:variations}. At the onset of TMT, the inner orbit comprises of two closely orbiting black holes with orbital properties $a_{\rm{in}} = 25 \ \rm{R}_{\odot}$ and $e_{\rm{in}} = 0.2$. The outer orbit has $a_{\rm{out}} = 200 \ \rm{R}_{\odot}$ and $e_{\rm{out}} = 0.3$. The masses of the components are $m_1=m_2=40 \ \rm{M}_{\odot}$ and $m_3=25 \ \rm{M}_{\odot}$, with a tertiary envelope mass of $16.7 \ \rm{M}_{\odot}$. We choose a gas density of $\rho=10^{-8}\rm{g\ cm^{-3}}$ and a mass transfer rate of $\dot{m}_3 = \dot{m}_{\rm{nuc}}$. To determine the nuclear timescale of the tertiary, we assign the initial ZAMS mass and luminosity of the tertiary to $m_{3,i}=27 \ \rm{M}_{\odot}$ and $8.5\times10^4 \ \rm{L}_{\odot}$, giving a timescale of $2.9\times10^5 \ \rm{yr}$ (Eq. \ref{eq:m_nuc}). The triple has a mutual inclination $i_{\rm{mut}}=115^\circ$, which implies a retrograde orbit of the tertiary around the inner binary.

\begin{figure*}
    \includegraphics[width=\textwidth]{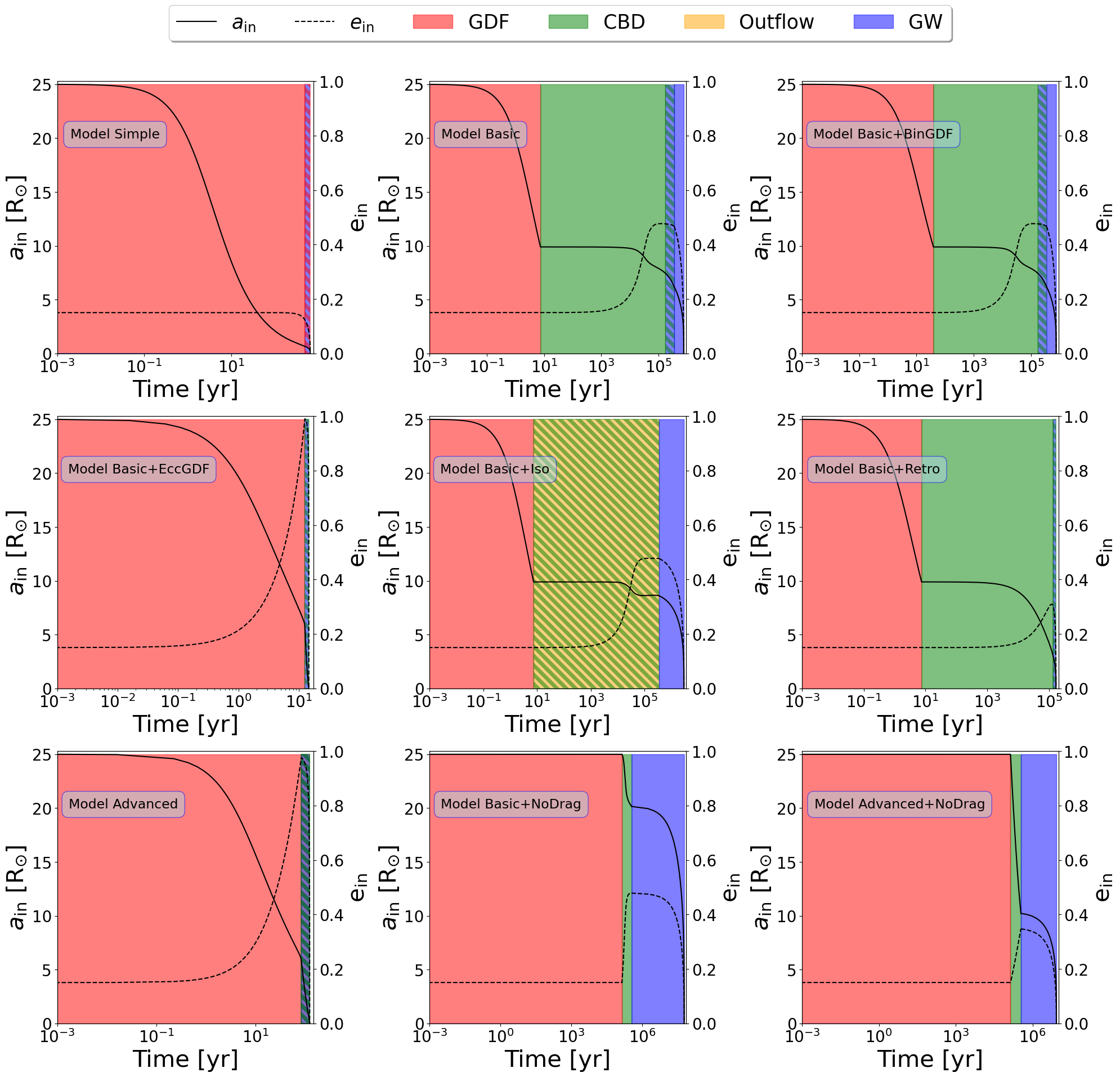}
    \caption{Evolution of the semi-major axis and eccentricity of the inner binary of an example system during the TMT phase and the subsequent GW inspiral. Different models are applied to the same system, as described in Section \ref{sec:variations}. The red shaded region indicates that the system undergoes ballistic accretion. The green shaded region indicates that the system has formed a CBD. The yellow region implies that the mass transfer is non-conservative during the CBD phase, and the material is assumed to be lost via an isotropic outflow. The blue shaded region indicates that the inspiral of the inner binary is dominated by GW emission. Overlapping colours show that more than one of the aforementioned processes apply.}
    \label{fig:inspiral_example}
\end{figure*}

In Fig. \ref{fig:inspiral_example}, we show the semi-major axis and eccentricity evolution of the inner binary during the TMT phase for all setups:
\begin{itemize}
    \item [-] \textbf{Model Simple:} we have assumed that the complete evolution is driven by BA, precluding CBD formation. The gas surrounding the binary exerts drag forces on the black holes, prompting a merger within 640 years. After 500 years, the gravitational wave inspiral starts to dominate over the ballistic accretion, resulting in a more rapid decay and circularisation of the orbit and an eventual merger.

    \item [-] \textbf{Model Basic:} the evolution of the system shows significant differences with Model Simple. For roughly the first 10 years, the inner orbit is too wide for a CBD to form with the current orbital configuration and the mass transfer proceeds in a ballistic fashion. During this period, the inner binary shrinks rapidly from $25 \ \rm{R}_{\odot}$ to $10 \ \rm{R}_{\odot}$. Afterwards, a CBD is formed and the evolution is directed by angular momentum exchange between the binary and the disk. As a consequence, during the next $1.7\times10^5$ years, the eccentricity increases quickly to the equilibrium value of about 0.48, while the orbit shrinks modestly, down to $7.5 \ \rm{R}_{\odot}$. For the following $1.8\times10^5$ years, angular momentum loss through GW emission becomes dominant, complementing the CBD torques. As a result, the eccentricity decreases, while the binary orbit shrinks more rapidly. The binary is not able to merge before the tertiary has transferred its entire envelope. At the termination of the TMT phase, the inner binary has a separation of $\rm{6.2 \ R_{\odot}}$. Subsequent GW emission brings the system to merger within another $4\times10^5$ years. Evidently, inspiral via BA is significantly more efficient than via CBD accretion for sufficiently high gas densities. It is noteworthy that a system that starts out in the regime of BA is always expected to transition to a CBD if initially $q_{\rm{out}} < 1$. The reason is that during BA, the outer orbit widens for the assumption of isotropic re-emission, while the inner orbit shrinks, facilitating the formation of a CBD.

    \item [-] \textbf{Model Basic+BinGDF:} the qualitative evolution of the inner orbit is similar to that of Model Basic. The only notable difference is that the transition from BA to CBD formation occurs later (after 40 years instead of 10 years). This is because the azimuthal drag force exerted by the wake of the companion is negative (it tends to speed up the object instead of slowing it down) and hence slowing the inspiral \citep{kim_dynamical_2008}. The final semi-major axis and eccentricity are close to identical as the transition to the CBD formation takes place at a similar inner separation. 

    \item [-] \textbf{Model Basic+EccGDF:} whereas Model Basic+BinGDF shows similar results to Model Basic, this model significantly deviates. During BA, the eccentricity of the inner binary increases drastically, reaching 0.99 after only 12 years. Because the apocenter of the inner binary increases at a faster rate than the semi-major axis decreases, the transition from BA to the formation of a CBD never occurs. Eventually, the eccentricity is sufficiently large, and the semi-major axis sufficiently small, for GW emission to dominate the evolution of the binary, leading to a rapid merger after 13.5 years. 

    \item [-] \textbf{Model Basic+Iso:} the evolution during the CBD phase is altered with respect to Model Basic. Since the conservativeness of mass transfer is only changed in the presence of a CBD, the evolution remains unchanged during BA. As the mass carried away has a specific angular momentum smaller than that of the binary orbit, the orbit effectively widens. The competing effects of the CBD torques, GW emission, and mass outflow leads to a slower decrease of the inner semi-major axis compared to Model Basic.  As a result, the separation at the termination of TMT is $\rm{8.6 \ R_{\odot}}$ (versus $\rm{7.5 \ R_{\odot}}$ for Model Basic).

    \item [-] \textbf{Model Basic+Retro:} similar to Model Basic+EccGDF, this model significantly deviates from Model Basic. During the CBD phase, the retrograde torques are more efficient at shrinking the orbit than for the prograde scenario, and the binary merges within $1.7\times10^5$ years. During the merger, the tertiary is still transferring mass towards the inner binary. During the final phase of evolution, the orbit is almost completely circularised by the GW emission.

    \item [-] \textbf{Model Advanced:} all features of the Basic+ models are combined. However, the evolution of the inner binary is very similar to that of Model Basic+EccGDF. This is because during BA the eccentricity rapidly increases, transitioning to GW dominated inspiral before a CBD can be formed. Therefore, the evolution of the inner binary only depends on the assumptions of the gas drag.

    \item [-] \textbf{Model Basic+NoDrag:} the inner binary remains unchanged during BA, as no drag forces are acting on its components. After approximately $10^5$ years the system transitions to a CBD phase. This shift occurs due to the outer orbit's expansion, caused by the redistribution and/or loss of angular momentum during the mass transfer process. The orbit shrinks to $\rm{20.2 \ R_{\odot}}$ once the mass transfer ceases, which is considerably wider than in models where drag forces are included. Nevertheless, GW emission leads to a merger within 60 Myr.

    \item [-] \textbf{Model Advanced+NoDrag:} the evolution follows a similar pattern to Model Basic+NoDrag, but during the CBD phase, retrograde torques cause the orbit to contract more efficiently to $\rm{10.2 \ R_{\odot}}$. As a result, the merger occurs significantly faster than in Model Basic+NoDrag, within 8.5 Myr. However, this is still at least a factor 20 longer than for models including GDF.

\end{itemize}

\section{Application to a triple population}
\label{sec:results_population}

Across both metallicities, we find that for about 7.5\% of triples in the simulated populations the inner binary evolves chemically homogeneous for the entire duration of the MS phase. For this subset of systems, the initial masses of the primary stars are at least $30 \ \rm{M_{\odot}}$. Among this population, 9.4\% of systems at $Z=0.005$ and 9.5\% at $Z=0.0005$ evolve towards TMT with an inner BBH. For these systems, we investigate the effect of TMT on their evolution and their prospects as GW sources. 

\subsection{Effect of TMT on inner orbits}
\label{sec:inner_orbits}

In this section, we examine the impact of TMT on the inner orbits consisting of a BBH, with a primary focus on the semi-major axis and eccentricity. We first discuss the population of high metallicity stars ($Z=0.005$), followed by a separate analysis of the role of metallicity.

\subsubsection{Inner semi-major axis}
\label{sec:evol_a}

\textbf{Only BA (Model Simple):} 
under the assumption that a CBD does not form, the semi-major axis at the end of TMT is mainly determined by the mass-transfer rate of the tertiary and the gas density of the cloud encompassing the inner binary during BA. A combination of high gas density and a long mass-transfer timescale results in very efficient inspiral of the binary, whereas low gas density and a short mass-transfer timescale lead to less efficient inspiral. This is evident from the final separations of the inner binary at the end of TMT: the average semi-major axis reduces from $49.0 \ \rm{R_{\odot}}$ to $34.4 \ \rm{R_{\odot}}$ under conditions of $\rho_g=10^{-10}\rm{ \ g\ cm^{-3}}$ with $\dot{m}_3 = \dot{m}_{\rm{th}}$. Conversely, all the inner orbits merge when $\rho_g=10^{-8}\rm{\ g\ cm^{-3}}$ with $\dot{m}_3 = \dot{m}_{\rm{nuc}}$.

\begin{figure}
    \includegraphics[width=\hsize]{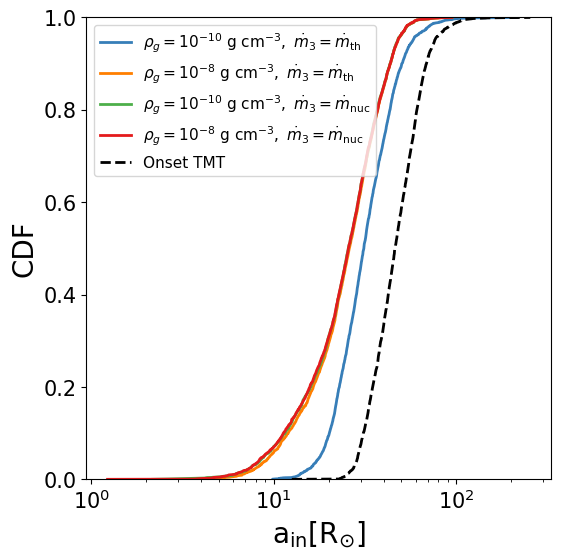}
    \caption{Cumulative distribution of the semi-major axis of the inner binaries at the end of TMT for Model Basic at $Z=0.005$. The few systems that have merged during TMT are omitted. The black dashed line denotes the semi-major axis distribution at the onset of mass transfer. Among different assumption for the gas density during BA and the mass transfer rate, the final distributions are fairly similar.}
    \label{fig:density_mt_rate}
\end{figure}

\textbf{BA \& CBD (Models Basic(+)/Advanced):}
if a CBD is allowed to form during TMT, the differences in the final semi-major axis between different assumptions for the gas density and mass transfer rate become less pronounced. In Fig. \ref{fig:density_mt_rate}, we show the inner semi-major axes after TMT for the four combinations of the mass transfer rate and gas density described in Sect. \ref{sec:variations}, applied to Model Basic. The average semi-major axes range from $26.7 \ \rm{R}_{\odot}$ to $34 \ \rm{R}_{\odot}$, indicating more consistent semi-major axes compared to scenarios considering only BA. Moreover, the final separations for all the combinations of gas density and mass-transfer rate, except for $\rho_g=10^{-10}\rm{g\ cm^{-3}}$ with $\dot{m}_3 = \dot{m}_{\rm{th}}$, are nearly identical. This implies that the final semi-major axis of the inner binary is not significantly dependent on the gas properties during BA, especially toward high gas densities. Moving forward, we will focus on two specific combinations of gas density and mass transfer rate for the other models that allow the formation of a CBD:  [$\rho_g=10^{-10}\rm{ \ g\ cm^{-3}}$ with $\dot{m}_3 = \dot{m}_{\rm{th}}$] and [$\rho_g=10^{-8}\rm{\ g\ cm^{-3}}$ with $\dot{m}_3 = \dot{m}_{\rm{nuc}}$].

Our findings indicate that the inclusion of eccentric drag forces during BA, the orientation of the CBD, and the conservativeness of mass transfer all have a significant influence on the evolution of the inner orbit. Systems that initially do not form a CBD can attain eccentricities close to unity, leading to an efficient merger through GW emission. However, if the orbit shrinks more rapidly than the eccentricity increases, a CBD might be formed before a high eccentricity is reached. Among the entire population, up to 20\% of the inner binaries are expected to merge during TMT due to the inclusion of eccentric drag forces. For systems that do not have a merger during TMT, the final inner semi-major axes are similar to Model Basic. Similarly, if we assume that the triples with a tertiary in a retrograde orbit relative to the inner binary can also form a retrograde CBD, the inner orbits at the end of TMT are more compact. On average, the inner semi-major axis is $12-13 \ \rm{R}_{\odot}$ smaller compared to scenarios considering only prograde disks. Isotropic mass outflow from the vicinity of the BHs on the other hand, leads to widening of the orbit through loss of angular momentum. If the TMT is completely non-conservative, the typical semi-major axis at the end of TMT is approximately $10 \ \rm{R_{\odot}}$ wider compared to completely conservative mass transfer. 

\textbf{No drag during BA (Models NoDrag):} neglecting drag forces entirely during BA leads to significantly wider final separations of the inner binaries. For systems initiating with BA, the average semi-major axes can be up to four times larger by the end of TMT compared to the models that include drag forces. However, in most systems, a CBD is expected to form immediately at the onset of the mass transfer, and its subsequent evolution remains consistent with models that account for drag forces.

\subsubsection{Eccentricity}

The evolution of the eccentricity of the inner binary is mainly governed by the interplay between the CBD and GW emission. For prograde disks, the eccentricity evolves towards an equilibrium point (until GW effects take over), as illustrated by the example system in Fig. \ref{fig:inspiral_example}. Consequently, for prograde disks, the typical inner eccentricities increase from 0.22 to 0.47-0.48 at the end of TMT, aligning with the equilibrium eccentricity for near-equal masses. We note that the semi-major axis consistently decreases once this equilibrium eccentricity is reached. This trend holds across the range of mass ratios considered in this study. However, it may not apply for significantly unequal mass ratios (see Fig. \ref{fig:cbd_deriv}). In contrast, systems influenced by retrograde disk torques do not reach an equilibrium eccentricity; instead, the eccentricity continually increases towards unity. This high eccentricity leads to rapid loss of orbital energy via GW emission, resulting in fast inspiral of the binary. As GWs also carry away angular momentum, the system does not maintain high eccentricities for long and eventually becomes nearly circularised before merging.

However, if we include eccentric drag forces, the evolution of the eccentricity is most prominent during BA and the GW dominated regime. As demonstrated in the previous section, the eccentricity can increase towards unity during BA for many systems, leading to an efficient merger

\subsubsection{Metallicity}
\label{sect:metallicity}

In this section, we discuss the role of metallicity on the final semi-major axis and eccentricity of the inner binaries. While the evolution of the system during TMT is not directly affected by metallicity in our model, the properties at the onset of TMT can be vastly different. 

\begin{figure*}
    \centering
    \begin{subfigure}{0.7\textwidth}
        \includegraphics[width=\textwidth]{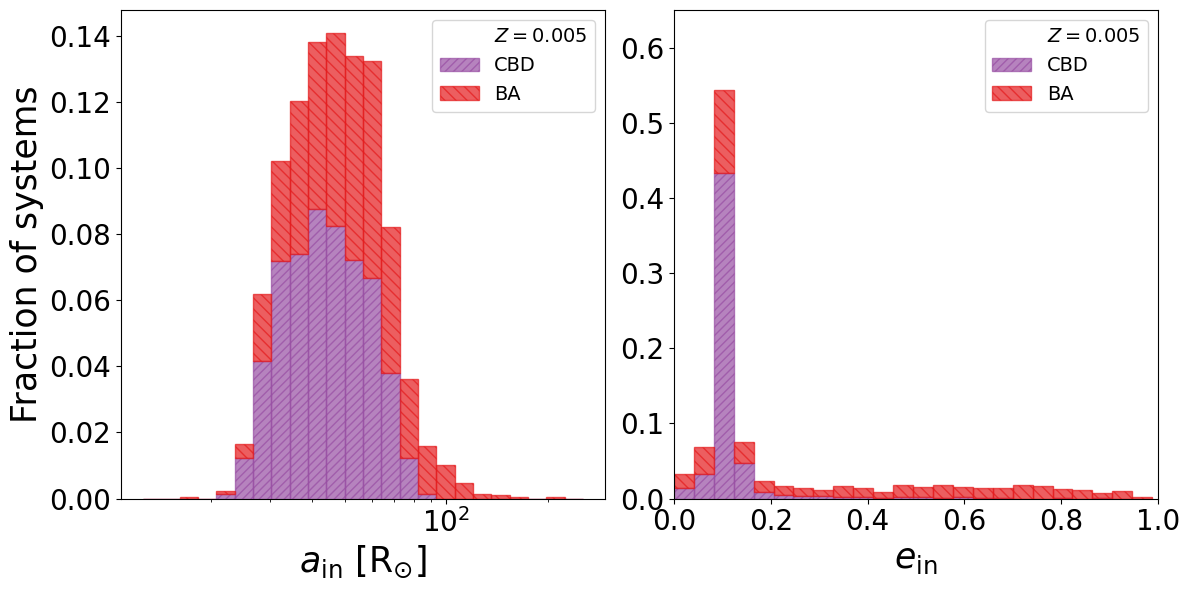}
        \label{fig:onset_a_e_highZ}
    \end{subfigure}
    \hfill
    \begin{subfigure}{0.7\textwidth}
        \includegraphics[width=\textwidth]{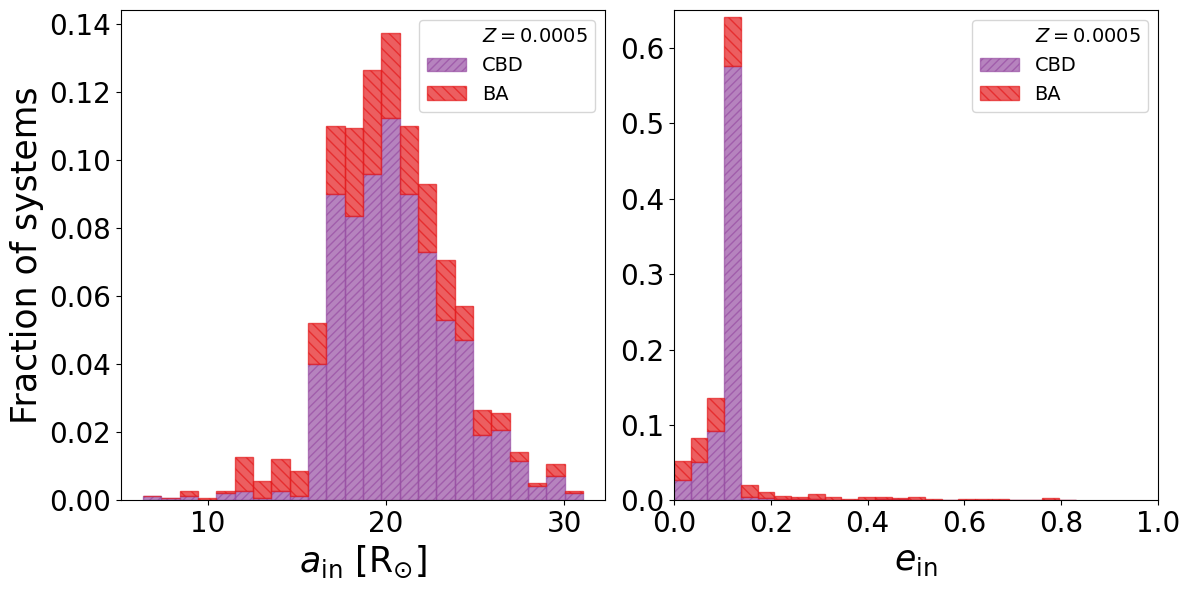}
        \label{fig:onset_a_e_lowZ}
    \end{subfigure}
    \caption{Distribution of the semi-major axis and eccentricity of the inner binaries with a metallicity of $Z=0.005$ (top) and $Z=0.0005$ (bottom) at the onset of TMT. The purple and red shaded regions correspond to systems that start the TMT phase with the formation of a CBD and ballistic accretion, respectively. The red bars are plotted on top of the purple bars. Note that the range of the semi-major axis is different between the upper and lower panel.}
    \label{fig:onset_a_e}
\end{figure*}

Firstly, stellar winds are metallicity dependent \citep{vink2001mass, vink2005metallicity} and effectively widen the inner orbit before the stars evolve into BHs, mainly during the post-MS phase of the BH progenitors. Therefore, high-metallicity systems are expected to develop wider orbits, whereas low-metallicity systems tend to remain more compact. The quantified effect of winds on the inner semi-major axes at the onset of TMT is shown in the left panels of Fig. \ref{fig:onset_a_e}. For systems with a metallicity of $Z=0.005$, the orbits range from approximately 20 to 200 $\rm{R}_{\odot}$, with an average of $49.0 \ \rm{R}_{\odot}$. Notably, the smallest inner semi-major axes are comparable to the MS radii of the BH progenitors, while the widest orbits significantly exceed the initial upper sampling limit of $40 \ \rm{R}_{\odot}$. In contrast, systems with a lower metallicity of $Z=0.0005$ experience considerably weaker stellar winds, resulting in a narrower range of semi-major axes at the onset of TMT. The widest orbits are approximately $30 \ \rm{R}_{\odot}$, with an average of $20.3 \ \rm{R}_{\odot}$. Some BBH systems have semi-major axes smaller than $15 \ \rm{R}_{\odot}$, which is less than the combined radii of their MS progenitors. These systems have been subjected to strong 3-body dynamical interactions, coupled with either tidal forces or GW emission, depending on the stellar type of the inner binary stars, resulting in energy dissipation from the binary. 

Secondly, the wider inner orbits of high-metallicity stars also lead to more pronounced 3-body dynamical interactions during the post-MS phase of the inner binary components compared to the lower metallicity systems. This is because the timescale of the ZLK cycles scales with the ratio between the outer and inner orbital period, and the separations of the outer orbits that undergo TMT are comparable across both metallicities considered in this study. Therefore, systems with metallicities of $Z=0.005$ have higher eccentricities at the onset of TMT than systems with metallicities of $Z=0.0005$, as shown in the right panels of Fig. \ref{fig:onset_a_e}. In both cases, the majority of systems have a non-zero eccentricity, with a prominent peak just above 0.1, attributed to the Blaauw kick experienced by BHs in our models due to neutrino losses. However, we predict a tail extending from eccentricities above 0.15 up to 1 only for the high-metallicity systems. 

Thirdly, the properties of the inner and outer orbits at the onset of TMT play a crucial role in determining the formation of a CBD. With wider and more eccentric inner orbits at higher metallicities, the likelihood of disk formation decreases. Initially, 57.1\% of systems with a metallicity of $Z=0.005$ form a CBD, whereas for systems with a metallicity of $Z=0.0005$, the percentage increases to 74.4\%.

\begin{figure}
    \includegraphics[width=\linewidth]{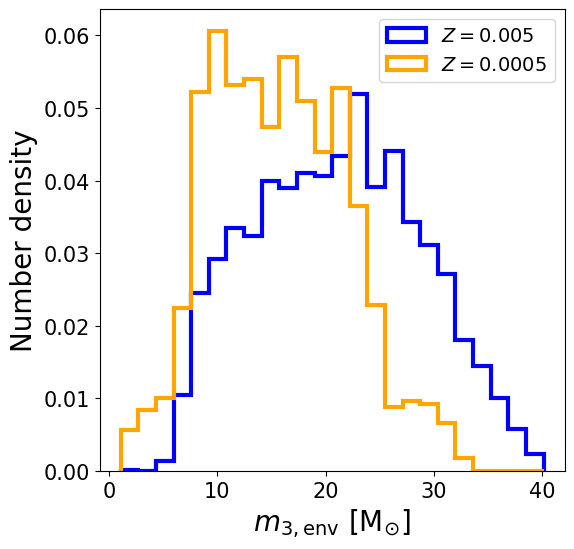}
    \caption{Distribution of envelope masses of the tertiary star at the onset of the TMT phase at metallicities of $Z=0.005$ and $Z=0.0005$.}
    \label{fig:m_env}
\end{figure}

Lastly, stellar winds determine the available envelope mass of the tertiary to be transferred during TMT. A higher metallicity, and thus higher mass-loss rate, would typically be expected to result in a smaller envelope mass of the donor star at the onset of TMT. However, our findings reveal the opposite, as shown in Fig. \ref{fig:m_env}. For systems with metallicities of $Z=0.005$, the mass ratios of the outer binary are typically larger than those in lower metallicity systems, as the BHs in the inner binary have lower masses due to higher wind mass loss from their progenitors. As a result, the Roche lobe of the tertiary is smaller, causing the star to fill the Roche lobe at an earlier evolutionary stage. Consequently, these stars frequently engage in mass transfer while still evolving through the Hertzsprung Gap (HG). Conversely, at $Z=0.0005$, the majority of tertiary stars are already supergiants, which are much more luminous and have experienced significantly stronger wind mass loss during the pre-TMT evolution. However, the mass-loss rates of massive stars near the Humphreys-Davidson limit that are classified as luminous blue variables (LBVs) are highly uncertain. Current models typically assume the mass-loss mechanism is metallicity independent, which may overestimate the mass lost at low metallicities.

The combined effects of the factors discussed above result in different final properties for the inner binary at the end of TMT. Specifically, in low metallicity systems, the typical inner semi-major axis ranges from $8.7 \ \rm{R}_{\odot}$ to $18.1 \ \rm{R}_{\odot}$, depending on the physical assumptions of the model. For the high metallicity systems, the separation is, on average, twice as large. Regarding the eccentricities, the distinction between the two populations is much less significant due to the CBD driving the systems towards the same equilibrium eccentricity.

\subsubsection{Effect of TMT on outer orbits}

Mass transfer from the tertiary to the inner binary inevitably leads to changes in the semi-major axis of the outer orbit due to redistribution or loss of orbital angular momentum. For all systems, TMT occurs in a stable manner (even for donor stars with a convective envelope), largely because of the prevalence of unequal mass ratios in the outer orbit. At metallicities of $Z=0.005$ and $Z=0.0005$, the mass ratio $q_{\rm{out}}$ does not exceed 1 and 0.6, respectively. If the mass ratio were higher, the tertiary would have initially been more massive than the stars of the inner binary, causing it to evolve and fill its Roche lobe before the inner binary became a BBH \citep{dorozsmai_stellar_2024}. Since the mass ratio is always less than 1, Eq. \ref{eq:da_out} indicates that the orbit widens throughout the entire mass transfer phase under the assumption of either conservative mass transfer or isotropic re-emission.

\subsection{GW sources}
\label{sec:gw_sources}
Having demonstrated that the inner binary can significantly shrink during TMT, we anticipate efficient formation of GW mergers. Assuming the absence of TMT, already a sizeable fraction of the population would be able to merge solely due to orbital inspiral via GW emission. Specifically, at $Z=0.005$ ($Z=0.0005$), 44.4\% (100\%) of the inner binaries are predicted to merge within a Hubble time. When TMT is included, we predict that 85.1-99.9\% of systems at $Z=0.005$ could form a GW merger within a Hubble time, assuming a CBD is formed. Without a disk, the merger fractions range between 91.5-100\%. 

Additionally, some of the mergers occur while TMT is still ongoing, potentially producing a GW signal accompanied by an EM counterpart. We discuss this further in Sect. \ref{sec:EM_signal}. We predict that $<0.03-46.8\%$ and $<0.07-29.3\%$ of systems could have an EM signal during the merger with metallicities of $Z=0.005$ and $Z=0.0005$, respectively. However, if we assume no CBD is formed, up to 100\% of systems can produce such mergers. The large uncertainty in these fractions arises from differences in metallicity, mass transfer rate, and physics of the BA and the CBD phase. An overview of all the merger fractions can be found in Table \ref{table:merger_fraction}.

\subsubsection{Eccentricities with aLIGO}

Binary eccentricity can serve as a key indicator to distinguish between different formation channels of GW mergers, as eccentric mergers are often linked to a dynamical history. Various dynamical pathways have been proposed to produce mergers with eccentricities exceeding 0.01-0.1 when entering the aLIGO frequency band at 10 Hz, such as isolated triple stars \citep{antonini_binary_2017, liu_black_2019}, galactic nuclei \citep{antonini_secular_2012, takatsy_eccentricity_2019, gondan_high_2021, samsing_agn_2022}, and cluster environments \citep[e.g.,][]{antonini2016black,samsing_eccentric_2018,rodriguez2018post1,rodriguez2018post2,antonini2020population,dallamico_eccentric_2024}. At design sensitivity, current ground-based GW detectors will be sensitive to eccentricities of at least 0.05 \citep{lower2018measuring}. Future third generation detectors are expected to be sensitive to eccentricities as low as $10^{-3}-10^{-4}$ \citep{lower2018measuring,saini2024resolving}. We explore whether eccentric BBH mergers can form via the evolutionary channel considered in this study. The GW frequency is determined according to the highest order amplitude \citep{wen_eccentricity_2003}:
\begin{equation}
    f_{\rm{GW}} = \frac{\sqrt{G(m_1+m_2)}}{\pi}\frac{(1+e)^{1.1954}}{[a(1-e^2)]^{1.5}}.
\end{equation}
We have shown that the interactions between the inner binary and the surrounding gas can increase the eccentricity and hence may counteract the circularisation of the orbit due to GW emission, thereby maintaining an eccentric orbit.

\begin{figure}
    \includegraphics[width=\linewidth]{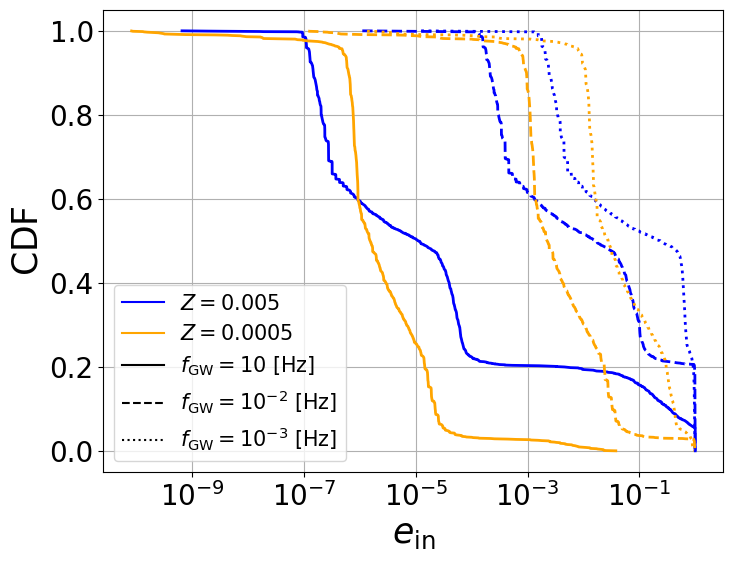}
    \caption{Cumulative eccentricity distribution for the GW mergers of Model Advanced with $\rho_g=10^{-8}\rm{\ g\ cm^{-3}}$ and $\dot{m}_3 = \dot{m}_{\rm{nuc}}$, when entering the aLIGO frequency band (solid lines) and in the frequency range at which LISA will be most sensitive (dashed and dotted lines).}
    \label{fig:ecc_ligo}
\end{figure}

As shown in Fig. \ref{fig:ecc_ligo}, our predictions indicate a wide range of eccentricities for systems that enter the aLIGO frequency band, with three distinct peaks identified in the distribution when including all additional physical phenomena considered in this study. The largest population, with low eccentricities ranging from $10^{-7}-10^{-6}$, consists of systems whose final phase of the spiral-in is driven purely by GW emission. This peak shows little dependence on metallicity and is at the same location also characteristic of GW mergers in field binaries and dynamical captures and ejections in globular clusters
\citep[e.g.,][]{nishizawa_constraining_2017,zevin2021implications}. The second peak, with eccentricities ranging between $10^{-5}-10^{-4}$, arises from systems with retrograde disks that have a merger during the CBD phase. The position of this peak is highly sensitive to the mass-transfer rate, as the time derivative of eccentricity from CBD torques depends on it. Therefore, if the mass transfer takes place on the thermal timescale, the peaks shifts to eccentricities ranging between $10^{-3}-10^{-2}$. A high eccentricity peak is predicted only for systems with metallicities of $Z=0.005$ and high gas densities, reaching eccentricities up to order unity. These systems originate from eccentric gas drag during BA. We find that 19\% of these systems reach eccentricities exceeding 0.01. However, we stress that the assumptions for the gas properties during BA are highly simplified and uncertain. When excluding drag forces altogether, the peak at the highest eccentricities disappears.

\subsubsection{Eccentricities with LISA}

Similar to aLIGO, future space-based GW detectors like LISA could provide valuable insights into the formation history of merging stellar-mass BHs \citep[see][]{amaro-seoane_astrophysics_2023}. Eccentricity is a key property in this context \citep{breivik_distinguishing_2016, nishizawa_constraining_2017, samsing_black_2018, zevin_eccentric_2019}, with predictions indicating that eccentricities of 0.001 should be measurable for 90\% of systems within a 5-year observation period, and for 25\% of systems within a 2-year period \citep{nishizawa_elisa_2016}. 

When the inner binaries enter the LISA frequency band at 0.001 Hz, they typically have semi-major axes of $1-1.5 \ \rm{R_{\odot}}$. By this stage, the systems have generally completed TMT, except for those mergers with a potential EM counterpart. Specifically, systems with an EM counterpart maintain high eccentricities throughout the LISA band because the circularisation due to GW emission is counteracted by CBD torques or eccentric gas drag, which drive the system to elliptical orbits. In Fig. \ref{fig:ecc_ligo}, we show the eccentricities at GW frequencies of 0.001 and 0.01 when including all variations of the physical phenomena. We predict similar peaks in the eccentricity distribution as for the aLIGO frequency band, spread across a wide range of eccentricities: $0.0001-1$. Consequently, the typical eccentricities can be characteristic of all types of evolutionary models, from mergers in isolated binaries to mergers through three-body interactions \citep[see][]{wang_space-based_2024}, depending on the physical processes involved in the merger. We note that even without gas drag, high eccentricities (e > 0.1) can be achieved, although it is less common by a factor two compared to cases where eccentric gas drag is included.

\subsubsection{Delay Times}

We define the delay time as the time between the formation of the triple and the merger of the compact objects:
\begin{equation}
    \tau_{\rm{delay}} = \tau_{\rm{tse}} + \tau_{\rm{TMT}} + \tau_{\rm{insp}},
\end{equation}
where $\tau_{\rm{tse}}$ is the time from the ZAMS up to the onset of TMT, $\tau_{\rm{TMT}}$ is the duration of the TMT, and $\tau_{\rm{insp}}$ is the inspiral time of the inner binary when its evolution is only driven by GW emission after the TMT phase. The delay time is an indicator for when a merger occurs and can be translated to a cosmic distance since GWs travel with the speed of light. Given the limited sensitivity of the current ground-based GW detectors, the distance of the merger affects its observability. 

\begin{figure}
    \includegraphics[width=\linewidth]{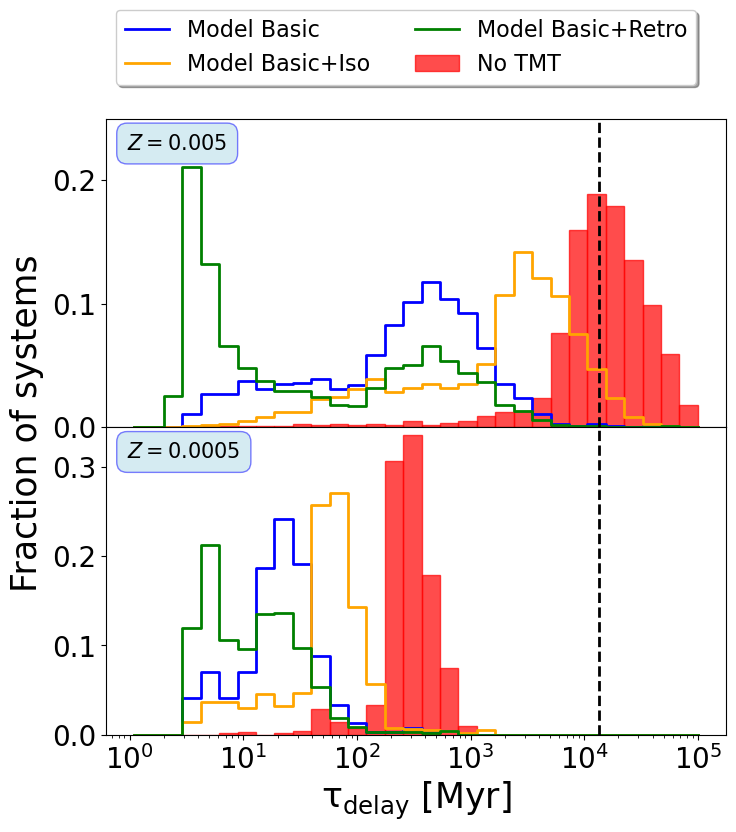}
    \caption{Delay time distribution of BBH mergers for different physical assumptions at $Z=0.005$ (top) and $Z=0.0005$ (bottom). The red histograms show the delay times, if TMT had not taken place. The black dashed line denotes the Hubble time of approximately 13.5 Gyr.}
    \label{fig:tdelay}
\end{figure}

Our findings show that the typical delay time shortens due to TMT, regardless of the physical assumptions, as demonstrated in Fig. \ref{fig:tdelay}. However, there is a considerable variation in the delay times across different models. Non-conservative mass transfer during the CBD phase leads to delay times around 3-6 times longer compared to conservative mass transfer.  Allowing for the formation of retrograde disks shortens delay times by 1.5-2 times compared to only prograde disks. Moreover, systems with low metallicity have significantly shorter delay times compared to high metallicity systems. As discussed in Sect. \ref{sect:metallicity}, the lower metallicity systems typically end up in more compact orbits at the end of TMT. Given the strong dependence of the GW inspiral timescale on the semi-major axis, the delay times at $Z=0.0005$ are 1 to 2 orders of magnitude shorter than at $Z=0.005$. We assess the impact of these differences in the delay times on the cosmic production rate of GW events in the next subsection.

\subsubsection{Merger Rates}

Even though the formation efficiency of GW sources is high for the studied channel, triples with a CHE inner binary can only be formed in a limited range of inner semi-major axes, stellar masses, inner mass ratios, and metallicities. We calculate the local merger rates of the inner binaries at redshift $z=0$. First, we determine the efficiency at which mergers are formed via this channel within the complete stellar population. This is expressed as:
\begin{equation}
    \epsilon_{\rm{eff}} = f_{\rm{CHE}}\frac{N_{\rm{merge}}}{N_{\rm{simulated}}},
\end{equation}
where $f_{\rm{CHE}}$ is the fraction of the simulated parameter space relative to the entire parameter space in which stars are born. $N_{\rm{merge}}$ is the number of GW mergers produced in our simulations, while $N_{\rm{simulated}}$ is the total number of triple systems simulated. Next, we translate this merger efficiency to a birth-rate density, which gives the number density of merging systems born per unit star-forming mass:
\begin{equation}
    \mathcal{R}_{\rm{birth}} = \frac{1}{\tilde{M}}\sum_{Z_i}\text{SFRd}(Z_i,z_{\text{form}})\epsilon_{\rm{eff}},
\end{equation}
where $\text{SFRd}(Z_i,z_{\rm{form}})$ is the star-formation rate density at the redshift at which the system formed following \citet{madau2014cosmic, madau2017radiation}, and $\tilde{M}$ is the average mass per system. $Z_i$ is the metallicity the system, while $z_{\rm{form}}$ is the formation redshift. Finally, we compute the merger rate by integrating the birth rate over the delay times:
\begin{equation}
    \mathcal{R}_{\rm{merge}} = \frac{1}{\tilde{M}}\sum_{Z_i}\int^{t_{\rm{merge}}}_{0}\text{SFRd}(Z_i,z_{\text{form}}(t_{\rm{delay}}))\epsilon_{\text{eff}}(\rm{t_{delay}})dt_{\rm{delay}},
\end{equation}
where $t_{\rm{merge}}$ is the time at which the merger occurs. For a more detailed description of how the merger rates were calculated, we refer to \citet{dominik2015double,dorozsmai_stellar_2024}.

The predicted local (redshift $z=0$) merger rates range between $0.69-1.74 \ \rm{Gpc}^{-3} \ \rm{yr}^{-1}$ depending on the assumed physics in the model. The variation in these rates is mainly a result of the different delay times for the mergers between the models. Interestingly, the high metallicity population contributes significantly more - by a factor 10-30 - toward the local merger rate than the low metallicity population. This discrepancy is due to the short delay times in the low metallicity population, which necessitates the formation of the original triples at a low redshift for them to be observable locally. However, at low redshifts, the star-formation rate (SFR) is relatively low, and stars tend to be born with higher metallicities, resulting in a low merger rate for this population. 

Regarding mergers with a potential EM counterpart, the local merger rate varies substantially across the different models. The predicted rates range from as high as $0.62 \ \rm{Gpc}^{-3} \ \rm{yr}^{-1}$ to almost negligible values. Models that predict higher rates typically include retrograde disks, eccentric gas drag, and high gas densities, as these conditions more efficiently facilitate GW mergers during TMT. However, observing these mergers locally is challenging because their delay times are generally short, often not exceeding a few tens of Myr. An overview of all merger rates can be found in Table \ref{table:merger_rates}.

\begin{figure}
    \includegraphics[width=\linewidth]{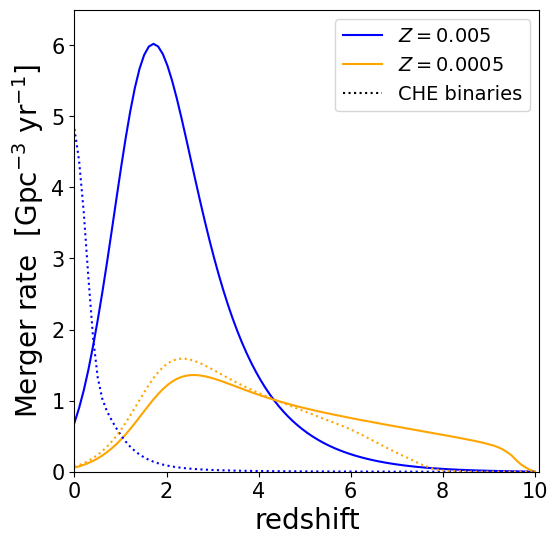}
    \caption{BBH merger rate as a function of redshift for Model Advanced at $Z=0.005$ (blue) and $Z=0.0005$ (orange). The dotted lines refer to the merger rate of triples with a CHE inner binary where the tertiary has not influenced the evolution of the inner binary.}
    \label{fig:rates_vs_z}
\end{figure}

Figure \ref{fig:rates_vs_z} illustrates the merger rates for both metallicity populations across redshift in our Model Advanced setup. The merger rate for both high and low metallicity populations increases steeply toward $z=2$, aligning closely with the peak of the cosmic star-formation rate. Beyond this point, the merger rate for the high metallicity population ($Z=0.005$) rapidly declines, while the merger rate for the low metallicity population ($Z=0.0005$) decreases more gradually. At redshifts greater than $z = 5$, most of the mergers originate from the low metallicity stars. This trend results from the combination of shorter delay times and the peak of the metallicity distribution shifting toward lower values at higher redshifts. 

Variations of the SFR and the cosmic metallicity distribution can significantly impact predictions for compact object merger rates \citep[e.g.,][]{neijssel_effect_2019, chruslinska_influence_2019, broekgaarden_impact_2021, broekgaarden_impact_2022}. This is especially true at high redshifts ($z>2$), where the SFR is not well established. Such uncertainty primarily affects local GW mergers from the high metallicity population in our study, as the short delay times of low metallicity mergers indicate that such mergers are only locally observable if the system was formed at a low redshift. Moreover, recent cosmological simulations of galaxy formation and evolution have identified a low metallicity tail at low redshifts \citep[e.g.,][]{van_son_locations_2023}, which is not captured by conventional log-normal distributions based solely on empirical evidence. While we do not quantify these uncertainties, they could significantly increase the merger rate for the low metallicity population.

Previous studies have already predicted redshift-dependent BBH merger rates for CHE binaries, but without a tertiary component \citep{mandel_merging_2016, riley_chemically_2021}. To compare with their findings, we computed the merger rate for systems where the tertiary star does not significantly influence the evolution of the inner binary, either through mass transfer or dynamical interactions, i.e., the inner binary evolves effectively isolated from the tertiary. As shown in Fig. \ref{fig:rates_vs_z}, the most pronounced difference with the studied triple channel is that the peak of the merger rate shifts towards lower redshift, particularly for the high metallicity population. This shift occurs because TMT generally shortens the merger delay time. \citet{mandel_merging_2016} considered a population at a fixed metallicity of Z=0.004, similar to our high metallicity population. Their merger rate peaks around a redshift of 0.3-0.4, somewhat higher than our findings, possibly due to minor differences in metallicity. Conversely, \citet{riley_chemically_2021} integrated the merger rate over a broader range of metallicities, using 30 evenly spaced metallicity bins, and found that the rate typically peaks at higher redshifts ($z=3-4$). This discrepancy however, might be attributed to differences in the sampled metallicity range, the number of different metallicities considered, SFR model, and redshift-dependent metallicity distribution of stars.  

\section{Discussion}
\label{sec:discussion}

In this section, we will address additional key observational signatures associated with GW mergers and examine the limitations of the mass transfer model.

\subsection{Gravitational wave sources}

\subsubsection{Electromagnetic signal}
\label{sec:EM_signal}
A merger of two BHs preceded (and accompanied) by strong EM emission would be a smoking gun to pin down the evolutionary origin of the merger. \citet{stone_assisted_2017} found that BBHs embedded in an AGN disk could be driven toward mergers assisted due to three-body interactions and interactions with the surrounding disk, potentially producing a luminous EM counterpart upon merger. Presumably, the X-ray emission is constrained by the Eddington mass accretion rate of the accreting object:
\begin{equation}
    L_{\rm{Edd}} = 1.3\times10^{38}\Bigg(\frac{M}{M_{\odot}}\Bigg) \ \rm{erg \ s^{-1}}.
\end{equation}
However, if the mass transfer rate exceeds the Eddington accretion rate significantly, the unaccreted mass is speculated to be expelled, obstructing the emitted X-ray radiation in most directions, thereby directing the emission towards the polar regions above and below the accretion disk \citep[e.g.,][]{king_ultraluminous_2023}. This leads to an observed X-ray luminosity in excess of $10^{39} \rm{erg \ s^{-1}}$ for neutron stars and stellar-mass black holes, classified as ultraluminous X-ray (ULX) sources. To date, the observed number of ULX candidates ranges in the thousands \citep{walton_multimission_2022}, with many originating from low-metallicity environments \citep{soria_star-forming_2005, mapelli_low_2009, prestwich_ultra-luminous_2013, basu-zych_exploring_2016, kovlakas_census_2020}. 

If the inner binaries of our triples manage to coalesce while the TMT phase is ongoing, this pathway could give rise to BBH mergers accompanied by EM signals. The BHs in our sample typically have Eddington mass accretion rates of $10^{-7}-10^{-6} \ \rm{M}_{\odot} \ \rm{yr}^{-1}$. The mass-transfer rate of massive tertiary stars, especially on the thermal timescale, far exceeds Eddington limit, leading to high X-ray luminosities. It is important to emphasize that the frequency of mergers with a disk still present heavily depends on the underlying model assumptions (see Sect. \ref{sec:gw_sources}).

\subsubsection{BH spin}

In this study we have not considered the spin distribution of merging BHs. The effective spin $\chi_{\rm{eff}}$ of the merging binary which contains both the magnitude and orientation of the spin vector for the binary system, is a property that can be inferred directly from the GW signal \citep{abbott_population_2023}. Disentangling the individual spin components poses a significant challenge. Previous studies of CHE in isolated binaries predict BH spins that stand out from other GW channels in terms of their high spin magnitudes retained from rapid rotation during the CHE phase \citep[e.g.,][]{zevin_one_2021}, although \citet{marchant_upper_2023} predict that the majority of locally detected BBHs from this channel have $\chi_{\rm{eff}} < 0.5$. In our sample, interactions between the disk and the binary could potentially alter the spin properties of the merging BHs. For instance, in inclined accretion disks, the BHs are driven to alignment by the Lense-Thirring effect \citep[e.g.,][]{volonteri_distribution_2005}, although the spins of the BHs are already expected to be aligned with respect to each other. An additional consequence of Lense-Thirring is the alignment between the angular momentum vector of the binary and the disk (and thus the tertiary), which could suppress dynamical interactions within the triple. This would prevent misalignment due to 3-body interactions during the successive inspiral of the binary. Furthermore, accretion of material could result in spin-up of the objects. However, \citet{munoz_circumbinary_2020} find that the time-averaged spin torques are only a few percent of the rate of change in orbital angular momentum. A more detailed investigations of the spin evolution in this channel is necessary to accurately determine the final spin distribution, but we anticipate the spins to be relatively similar to CHE in isolated binaries.

\subsubsection{Pulsational pair-instability supernovae}

Stellar rotation decreases the initial mass necessary to form helium and CO cores of a given mass \citep{chatzopoulos_2012}, an effect that is particularly relevant for stars that spin so fast that they evolve chemically homogeneous \citep{woosley_pulsational_2017}. For helium and CO core masses of about $30\,\text{M}_{\odot}$ or more this leads to pulsational instabilities in the very final phases of evolution. These may lead to very substantial mass shedding, reducing the mass and angular momentum of the black hole that forms in the (pulsational pair-instability) supernova that ensues \citep[see also,][]{marchant_pulsational_2019, stevenson_impact_2019}. 

For the explored parameter space (i.e., initial masses of the primary component of the inner binary between 20 and 100\,$\text{M}_{\odot}$) 7.5\% of systems evolves chemically homogeneous, out of which 9.5\% lead to mass transfer from the tertiary to an inner BBH. From these, for $Z = 0.005$, 50\% of the inner binary components ultimately have helium and CO core masses in the range $35-60 \ \rm{M}_{\odot}$. For $Z=0.0005$ this is 99\%. So, the majority of our systems may be affected by pulsational pair instability effects not accounted for in our simulations. The primary effect the pair instability driven mass loss is anticipated to have is that it drives the inner binary toward larger separations. 
The implications of this shift on the merger process remain ambiguous. On the one hand, the inspiral time should increase as the binary widens. On the other hand, three-body dynamical interactions become stronger, leading to more eccentric orbits. Additionally, the final BH masses and angular momenta would be reduced. The former influences the evolution of both the inner and outer orbits throughout the entire TMT phase and subsequent GW inspiral. Furthermore, the mass transfer is more prone to be unstable.

\subsection{Assumptions regarding circumbinary disks}

\subsubsection{Viscosity parameter}

The evolution of the inner binary in the presence of a CBD in our model are based on simulations taylored for SMBHs \citep{siwek_orbital_2023}, though our investigation primarily concerns stellar mass objects, which are significantly lower in mass. Notably, the viscosity parameter $\alpha$ in disks around stellar mass systems are typically around an order of magnitude lower than those around SMBHs \citep{hartmann_accretion_1998}. \citet{dittmann_survey_2022} found that for low aspect ratios, $H/r < 0.1$, the transfer rate of specific angular momentum between the disk and the binary is highly dependent on the viscosity of the disk. However, at $H/r = 0.1$, this dependency more or less disappears, resulting in consistent semi-major axis evolution regardless of the disk's viscosity. This result is in agreement with the findings of \citet{munoz_circumbinary_2020}. Given that stellar-mass disks typically have aspect ratios closer to 0.1 \citep[see][]{lai_circumbinary_2022}, it suggests that the binary's evolution may not deviate significantly.

\subsubsection{Steady-state disk}

In our simulations we have assumed that a steady state disk is reached instantaneously, wherein the mass accretion rate onto the inner binary matches the supply rate at the outer boundary of the disk. However, this assumption overlooks the so-called 'transient phase', characterised by the cavity's filling within the circumbinary disk and the formation of circumstellar disks encircling the individual binary components. This phase may involve loss of angular momentum from the inner binary as the net torque is provided solely by gravitational interactions \citep{munoz_circumbinary_2020}, leading to an additional decrease in the semi-major axis. Although the duration of the transient phase is dependent on the initial disk properties, we perform a back-of-the-envelope calculation to assess the significance of this phase. Following \citet{munoz_circumbinary_2020}, if a steady state is reached after a few hundred binary orbits, its duration could extend to approximately 10 years. However, for stronger viscous disks, resembling disks around stellar-mass objects instead of SMBHs, the transient phase is anticipated to last even longer \citep[e.g.,][]{miranda_viscous_2017}. Considering the thermal timescale of massive evolved stars, estimated at about 10-100 years, it becomes apparent that a significant portion of the mass transfer phase could be governed by the transient phase, and therefore the orbital shrinkage could be underestimated in the current model.

Similarly, we have assumed that the disk promptly disperses as soon as the mass-transfer phase is terminated. However, for a finite disk in quasi-steady state (where the mass-accretion rate scales with the rate at which the disk viscously spreads), the mass accretion onto the binary diminishes with time, following a time dependency of approximately $\dot{m}_{\rm{acc}} \propto t^{-4/3}$ \citep{munoz_circumbinary_2020, siwek_preferential_2023}. As a result, it is conceivable that our estimation of the disk's lifetime is underestimated. Increased disk lifetimes coupled with strong GW emission, may lead to an increase in the number of mergers in the presence of a CBD, consequently leading to a higher occurrence of mergers accompanied by an EM signal. This could be particularly pronounced at lower metallicity, where delay times can be less than 10 Myr, and GW emission plays a significant role in the orbital evolution. This scenario is especially relevant for massive post-MS tertiary stars, as their evolutionary timescale typically spans 0.1-1 Myr for the tertiary stars considered in this study. 

\subsubsection{Comparison with other studies}

Similar to our approach, \citet{valli_long-term_2024} used an interpolation method to analyse changes in semi-major axis and eccentricity, referencing the work of \citet{siwek_orbital_2023}, and compared these results with those from \citet{zrake_equilibrium_2021} and \citet{dorazio_orbital_2021}. Notably, disagreements arise regarding the stability of the eccentricity in initially circular orbits with equal mass ratios. While \citet{siwek_orbital_2023} suggests the configuration is unstable, leading to an increase in eccentricity and ultimately a decreasing orbital separation, the latter two studies indicate an additional equilibrium eccentricity exists at $e=0$ characterised by orbital widening. At the onset of TMT, fewer than 1\% (2.5\%) of the inner binaries have eccentricities below 0.01 for the $Z=0.005$ ($Z=0.0005$) population. Predominantly, inner binaries at both metallicities have eccentricities around 0.1 upon TMT onset, largely attributed to the Blaauw kick during BH formation. However, we assume 10\% of the mass is lost as neutrinos, which might be an overestimation \citep{vigna2024constraints} and lead to an overestimation of the number of eccentric orbits. Nevertheless, many triples are influenced by three-body dynamical effects, increasing the eccentricity regardless of the occurrence of a supernova kick. Therefore, the potential equilibrium eccentricity at $e=0$ is not of great importance to our systems.

\subsection{Assumptions regarding ballistic accretion}

\subsubsection{Properties of gas around the inner binary}

It is important to recognise that uncertainties related to the gas properties could lead to considerable variations in the evolution of the binary during TMT, potentially influencing the findings presented in this study.

For the duration of the BA phase we relied on simplified assumptions regarding the gas enveloping the inner binary. These simplifications stem from our limited knowledge, largely due to the scarcity of detailed studies into stable mass transfer from tertiary stars. To encompass a wide range, we chose two significantly different gas densities ($\rm{\rho_{gas}=10^{-8}g\ cm^{-3}}$ and $\rm{\rho_{gas}=10^{-10}g\ cm^{-3}}$), which are in reasonable agreement with densities observed in previous research \citep{de_vries_evolution_2014}. Additionally, we imposed a constraint that the density remains several orders of magnitude lower than that of a typical common envelope structure. 
In Sect. \ref{sec:inner_orbits}, we demonstrated that the final orbital separations of the inner binary at the end of TMT are more similar when considering the formation of a CBD, even for different 
densities of the gas during BA.

Additionally, our model assumes that the density of gas surrounding the inner binary remains constant during the BA phase. This assumption implies a balance between material ejected from the inner binary and mass transferred from the tertiary. If we assume the gas density increases in time, starting from a high initial gas density, the final properties of the inner binary are not expected to change much, as these are mainly determined by the CBD phase. Conversely, starting from a low initial gas density, the final orbital properties might be different if most or the entire mass transfer takes place during BA. Alternatively, we consider a scenario where the gas density around the inner binary decreases over time. This leads to depletion of the gas cloud, potentially completely halting the inspiral. However, the cloud continuously receives mass from the tertiary, maintaining some gas around the binary during the TMT phase. Characterising the properties and evolution of such gas clouds would require more detailed studies of stable TMT.

Another assumption we have made is that the density of the gas is completely homogeneous and has no dependence on the distance from the COM, resulting in efficient inspiral of the binary. \citet{antoni_evolution_2019} argue that if the COM is at rest relative to the gas, the density near the COM becomes enhanced, leading to a Bondi density profile where $\rho_g \propto a^{-3/2}$. Consequently, the inspiral becomes more efficient for supersonic motion and less efficient for subsonic motion. Since we expect the compact objects in our study to typically move at supersonic velocities,
the transition time from BA to a CBD would become shorter. Furthermore, a centrally concentrated mass distribution would lead to a less efficient increase in eccentricity, or even a decrease for a sufficient high radial dependence of the gas density  \citep{szolgyen_eccentricity_2022}. This would affect the subsequent CBD evolution, but its impact on merger frequency and properties is not evident.

The gravitational drag force experienced by binary objects is significantly influenced by the relative velocity between the gas and the inspiraling objects. This dependence arises because $F_{\rm{drag}} \propto v_{\rm{rel}}^{-2}$ for supersonic motion relative to the gas. This relation originates from the accretion radius of the binary objects, which shares the same proportionality with the relative velocity. In our simulations, we have assumed the gas to be at rest with respect to the COM of the inner binary. This is not a realistic assumption as the gas stream enters the vicinity of the inner binary with an initial velocity relative to the COM and is subsequently disturbed upon intersecting with the binary orbit. A prograde (retrograde) direction of the gas velocity would lead to more (less) efficient binary inspiral during BA. 

The findings of \citet{ostriker_dynamical_1999,kim_dynamical_2007} on the GDF were derived under the assumption of an infinite background. In our scenario, the binary is embedded within a finite amount of gas during BA. A back-of-the-envelope calculation suggests that, given the gas densities under consideration, the total mass of the surrounding gas cloud is likely much lower than the mass of the inner binary components. Although the precise impact of this difference on the results of \citeauthor{ostriker_dynamical_1999,kim_dynamical_2007} is unclear, it is expected that the efficiency of GDF is reduced, resulting in a less effective inspiral of the binary

\subsubsection{Hydrodynamic drag during ballistic accretion}

Computing the orbital evolution of the binary embedded in a gaseous medium in our model, we have only considered the effect of the gravitational drag force that arises through the interaction between the overdense wake and the orbiting bodies. Additional drag forces can be imparted by direct impact of the gas on the bodies, resulting in a loss of angular momentum. This hydrodynamic drag can be expressed as:
\begin{equation}
    \boldsymbol{F}_{\rm{drag}} = -\frac{1}{2}C_{D}\pi R^2\rho_{g}v_{\rm{rel}}\mathbf{v_{\rm{rel}}},
\end{equation}
where $C_D$ is the drag coefficient. 
The balance between both types of drag force can be estimated in terms of the compactness, $M/R$, of the objects \citep{grishin_application_2015, szolgyen_eccentricity_2022}. If the compactness exceeds a critical value, the gravitational drag force dominates over the hydrodynamic drag force, and vice versa. In this study, we applied the TMT model to inner binary objects consisting of BHs, where we found the hydrodynamic drag force to be negligible as the accretion radius of the BHs, $R_\text{acc} = 2Gm/v_{\text{rel}}^2$, is significantly larger than their Schwarzschild radius. Therefore, considering only hydrodynamic drag we find a similar evolution of the systems as when drag is completely neglected. However, if the mass transfer phase is initiated while the stars of the inner orbit are still on the main sequence, as can happen when the tertiary star is initially the most massive object in the triple \citep[e.g.,][]{toonen_evolution_2020}, the contribution of the hydrodynamic drag is not evident. 

\begin{figure}
    \includegraphics[width=\linewidth]{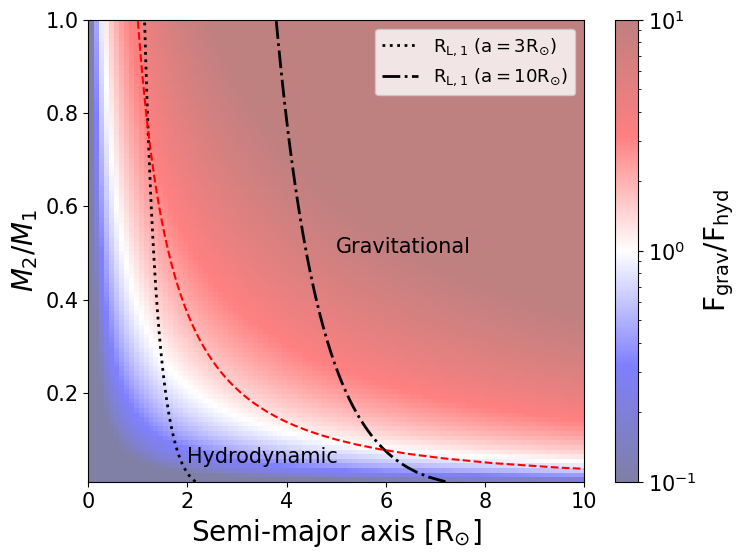}
    \caption{Dominant type of drag force (gravitational versus hydrodynamic) as a function of the semi-major axis and the mass ratio for a secondary with $M_2 = 1 \ \rm{M}_{\odot}$. The red dashed line indicates the radius of the primary ($R_1$) at the ZAMS. The dotted and dash-dotted lines refer to the Roche radii of the primary at a semi-major axis of $3 \ \rm{R}_{\odot}$ and $8 \ \rm{R}_{\odot}$, respectively. If the radius exceeds the size of the Roche lobe, the system should not exist.}
    \label{fig:compactness}
\end{figure}

We estimate the parameter space where the hydrodynamic drag is important. Ignoring the constants that are approximately equal to one, the gravitational drag is dominant for each of the components if the following condition is satisfied \citep{szolgyen_eccentricity_2022}:
\begin{equation}
    \frac{M_i}{R_i} \gtrsim \frac{v_{i,\rm{rel}}^2}{2G}.
\end{equation}
For MS stars, the compactness increases toward larger masses, as the mass and radius relate as approximately $R \propto M^{0.7}$. Furthermore, the orbital velocity of the more massive component in a binary is lower than the velocity of its companion. Therefore, we only compare the drag forces for the secondary star. If the relative velocity is equal to the Keplerian velocity, we can express the equation in terms of semi-major axis and binary mass:
\begin{equation}
    \frac{M_2}{R_2} \gtrsim \frac{M_{\rm{bin}}}{2a\Big(1+\frac{M_2}{M_1}\Big)^2}.
\end{equation}
We consider a $1 \ \rm{M}_{\odot}$ secondary and calculate the ratio between the competing drag forces for a grid of mass ratios in the range 0.01-1 and semi-major axes in the range $0.01-10 \ \rm{R_{\odot}}$. We find that the hydrodynamical drag force becomes important toward unequal mass ratios and small separations, as shown in Fig. \ref{fig:compactness}. However, when only considering detached systems, the hydrodynamical drag is non-negligible only at separations of at least about $10 \ \rm{R}_{\odot}$ and mass ratios close to 0.1. For any other orbital configurations, the gravitational drag force dominates the total drag. However, if gravitational drag is absent, hydrodynamic drag could still lead to changes in the orbital properties. If the gas were to move move prograde (retrograde) with respect to the binary, the hydrodynamic drag becomes less (more) important. 

\section{Conclusions}
\label{sec:conclusion}

We presented a rapid analytical model designed to probe the evolution of triple star systems undergoing stable mass transfer from the tertiary component to the inner binary. This model serves as a tool for conducting large-scale population simulations. Our analysis describes two primary modes of mass transfer evolution: ballistic accretion, characterised by the inner binary engulfed within a gaseous cloud and governed by gravitational drag forces, and circumbinary disk  evolution, wherein gravitational and accretion torques dictate orbital changes. We applied this model to a simulated population of triples with a Roche-lobe filling tertiary star transferring mass onto an inner binary consisting of two black holes, at metallicities of $Z=0.005$ and $Z=0.0005$. Both components of the inner binary experienced chemically homogeneous evolution throughout their main-sequence phase. We summarise our key findings:

\begin{itemize}
    \item For triples with a CHE inner binary, tertiary mass transfer is typically stable due to the mass of the tertiary being lower than the mass of the inner binary. Among the simulated systems, 42.9\% (25.6\%) are not able to form a CBD at the onset of TMT at metallicities of $Z=0.005$ ($Z=0.0005$), mainly due to small ratios between the outer and inner orbital separation, and start out the evolution with BA. 
    \item The evolution of the inner binary during BA invariably leads to a decrease in the semi-major axis of the inner binary on a timescale much shorter than that of the CBD evolution if gas drag is efficient. Additionally, for initially eccentric inner binaries, drag forces can rapidly increase the eccentricity toward unity, leading to efficient formation of GW mergers. CBD evolution tends to shrink the orbit for the simulated range of near-equal mass ratio binaries. However, if the mass transfer is completely non-conservative, it may lead to widening of the inner orbit during this phase. Eccentricity typically converges towards an equilibrium value of approximately 0.48 under CBD interactions in prograde orbits. However, in cases where retrograde disks are formed, the eccentricity increases toward unity.
    \item While many systems start out with BA, the evolution of the orbital properties typically leads to the formation of a CBD. Whether a CBD can form primarily depends on the balance between the rate at which the inner semi-major axis shrinks and the outer orbit expands (promoting CBD formation) versus the rate at which the inner eccentricity increases (hindering CBD formation). If a CBD does form, the importance of the efficiency of the binary inspiral during BA — dictated by the properties of the gas surrounding the inner binary — diminishes, as the CBD regulates the inspiral.
    \item Even though the formation efficiency of systems that evolve through the evolutionary channel investigated in this study is low ($<1\%$ for systems with initial masses of the primary component of the inner binary between 20 and 100\,$\text{M}_{\odot}$ and separations of the outer orbit up to $10^5\,\text{R}_{\odot}$), the channel is highly efficient in producing GW mergers. For systems with a metallicity of $Z=0.005$, we predict that 85.1-100\% of inner binaries will merge within a Hubble time, depending on the specific physical model assumptions. At a metallicity of $Z=0.0005$, the efficiency reaches 100\%. This higher efficiency at $Z=0.0005$ is primarily due to minimal binary widening caused by stellar winds during pre-supernova evolution, resulting in inner orbits that are, on average, $30 \ \rm{R}_{\odot}$ less wide compared to systems with a metallicity of $Z=0.005$.
    \item The total local merger rate spans a range of 0.69-1.74 $\rm{Gpc^{-3} \ yr^{-1}}$, with the primary contribution originating from higher metallicity systems, despite their lower efficiency in forming GW mergers. This behaviour arises due to a combination of factors including the delay time distribution and the metallicity-specific star-formation rate. Notably, low-metallicity binaries have delay times only up to a few hundred Myr, suggesting that they predominantly merge at high redshifts, where the formation of low metallicity stars is favoured.
    \item Our predictions suggest there are two types of GW mergers: those occurring during TMT, and those taking place after the TMT phase has finished. The former type of mergers has the potential to produce an EM signal before and during the merger event. Although, the occurrence varies significantly depending on the model assumptions. Considering only prograde disks and ignoring eccentric gas drag, we predict an EM signal in up to 5.6\% of mergers at low metallicity, whereas including retrograde disks and eccentric gas drag could yield rates as high as 46.8\%. Even when ignoring gas drag completely during BA, we predict up to 19.2\% of mergers occurring during TMT. However, these rates are subject to large uncertainties, mainly due to the assumptions made for the initial triple orientations and the gas properties during BA.
    \item Our study reveals that triples with CHE inner binaries share several properties with other formation pathways, but also present distinct differences. Compared to CHE in isolated binaries, these systems are expected to have similar mass ratios and effective spins. However, unlike the isolated binary case, our findings indicate that systems with high eccentricities can be formed in the aLIGO and LISA frequency band, under the assumption that retrograde disks and eccentric gas drag play a role in the evolution of the population. Unlike dynamical formation pathways involving classically evolving triples or dynamical interactions in clusters, the effective spin in triples with a CHE inner binary is expected to be typically non-zero. Nevertheless, for most systems, the predicted eccentricities closely align with those found in and around cluster environments.

\end{itemize}

\begin{acknowledgements}
   FK would like to thank Alejandro Vigna-G{\'o}mez, Magdalena Siwek, Mor Rozner, Morgan Mcleod and Nathalie Degenaar for the helpful discussions. The authors also acknowledge support from the Netherlands Research Council NWO (VENI 639.041.645 and VIDI 203.061 grants). The data and scripts necessary to run the model and reproduce the figures in this paper are publicly available on Zenodo: \href{https://zenodo.org/record/13610538}{10.5281/zenodo.13610538}.
\end{acknowledgements}

\bibliographystyle{aa} 
\bibliography{ref} 

\begin{appendix}

\onecolumn
\section{Overview merger rates}
\begin{table}[h!]
\caption{Percentage of systems with TMT investigated in this study that have a BBH merger within a Hubble time. Without considering the effects of TMT, 44.4\% and 100\% of the systems would produce a BBH merger at metallicities of $Z=0.005$ and $Z=0.0005$, respectively.}           
\label{table:merger_fraction}  
\centering                         
\resizebox{0.8\textwidth}{!}{\begin{tabular}{l c c c}        
\hline\hline  
& Mergers during TMT (\%) & Mergers after TMT (\%) & Total Mergers (\%) \\
\hline\hline
  \textbf{Model Simple} \\
  $Z=0.005$, $\rho_g=10^{-8}$, $\dot{m}_3=\dot{m}_{\rm{nuc}}$  & 100 & <0.03 & 100 \\ 
  $Z=0.005$, $\rho_g=10^{-8}$, $\dot{m}_3=\dot{m}_{\rm{th}}$ & 0.4 & 99.6 & 100 \\
  $Z=0.005$, $\rho_g=10^{-10}$, $\dot{m}_3=\dot{m}_{\rm{nuc}}$ & 99.7 & 0.3 & 100 \\
  $Z=0.005$, $\rho_g=10^{-10}$, $\dot{m}_3=\dot{m}_{\rm{th}}$ & <0.03 & 91.5 & 91.5 \\
  $Z=0.0005$, $\rho_g=10^{-8}$, $\dot{m}_3=\dot{m}_{\rm{nuc}}$ & 100 & <0.07 & 100 \\
  $Z=0.0005$, $\rho_g=10^{-8}$, $\dot{m}_3=\dot{m}_{\rm{th}}$ & 4.8 & 95.2 & 100 \\
  $Z=0.0005$, $\rho_g=10^{-10}$, $\dot{m}_3=\dot{m}_{\rm{nuc}}$ & 99.6 & 0.4 & 100 \\
  $Z=0.0005$, $\rho_g=10^{-10}$, $\dot{m}_3=\dot{m}_{\rm{th}}$ & <0.07 & 100 & 100 \\ \hline
  \textbf{Model Basic} \\
  $Z=0.005$, $\rho_g=10^{-8}$, $\dot{m}_3=\dot{m}_{\rm{nuc}}$  & 0.4 & 99.2 & 99.6 \\ 
  $Z=0.005$, $\rho_g=10^{-8}$, $\dot{m}_3=\dot{m}_{\rm{th}}$ & <0.03 & 99.6 & 99.6 \\
  $Z=0.005$, $\rho_g=10^{-10}$, $\dot{m}_3=\dot{m}_{\rm{nuc}}$ & 0.2 & 99.4 & 99.6 \\
  $Z=0.005$, $\rho_g=10^{-10}$, $\dot{m}_3=\dot{m}_{\rm{th}}$ & <0.03 & 97.6 & 97.6 \\
  $Z=0.0005$, $\rho_g=10^{-8}$, $\dot{m}_3=\dot{m}_{\rm{nuc}}$ & 5.6 & 94.4 & 100 \\
  $Z=0.0005$, $\rho_g=10^{-8}$, $\dot{m}_3=\dot{m}_{\rm{th}}$ & <0.07 & 100 & 100 \\
  $Z=0.0005$, $\rho_g=10^{-10}$, $\dot{m}_3=\dot{m}_{\rm{nuc}}$ & 5.4 & 94.6 & 100 \\
  $Z=0.0005$, $\rho_g=10^{-10}$, $\dot{m}_3=\dot{m}_{\rm{th}}$ & <0.07 & 100 & 100 \\ \hline
  \textbf{Model Basic+BinGDF} \\
  $Z=0.005$, $\rho_g=10^{-8}$, $\dot{m}_3=\dot{m}_{\rm{nuc}}$ & 0.2 & 99.4 & 99.6 \\ 
  $Z=0.005$, $\rho_g=10^{-10}$, $\dot{m}_3=\dot{m}_{\rm{th}}$ & <0.03 & 94.4 & 94.4 \\
  $Z=0.0005$, $\rho_g=10^{-8}$, $\dot{m}_3=\dot{m}_{\rm{nuc}}$ & 4.9 & 95.1 & 100 \\
  $Z=0.0005$, $\rho_g=10^{-10}$, $\dot{m}_3=\dot{m}_{\rm{th}}$ & <0.07 & 100 & 100 \\ \hline
    \textbf{Model Basic+EccGDF} \\
  $Z=0.005$, $\rho_g=10^{-8}$, $\dot{m}_3=\dot{m}_{\rm{nuc}}$  & 20.3 & 79.3 & 99.6 \\ 
  $Z=0.005$, $\rho_g=10^{-10}$, $\dot{m}_3=\dot{m}_{\rm{th}}$ & 11.5 & 87.9 & 99.4  \\
  $Z=0.0005$, $\rho_g=10^{-8}$, $\dot{m}_3=\dot{m}_{\rm{nuc}}$ & 8.7 & 91.3 & 100 \\
  $Z=0.0005$, $\rho_g=10^{-10}$, $\dot{m}_3=\dot{m}_{\rm{th}}$ & 0.2 & 99.8 & 100 \\ \hline
    \textbf{Model Basic+Iso} \\
  $Z=0.005$, $\rho_g=10^{-8}$, $\dot{m}_3=\dot{m}_{\rm{nuc}}$ & 0.1 & 88.9 & 89 \\ 
  $Z=0.005$, $\rho_g=10^{-10}$, $\dot{m}_3=\dot{m}_{\rm{th}}$ & <0.03 & 85.1 & 85.1 \\
  $Z=0.0005$, $\rho_g=10^{-8}$, $\dot{m}_3=\dot{m}_{\rm{nuc}}$ & 2.3 & 97.7 & 100 \\
  $Z=0.0005$, $\rho_g=10^{-10}$, $\dot{m}_3=\dot{m}_{\rm{th}}$ & <0.07 & 100 & 100 \\ \hline
    \textbf{Model Basic+Retro} \\
  $Z=0.005$, $\rho_g=10^{-8}$, $\dot{m}_3=\dot{m}_{\rm{nuc}}$ & 37.1 & 62.8 & 99.9 \\ 
  $Z=0.005$, $\rho_g=10^{-10}$, $\dot{m}_3=\dot{m}_{\rm{th}}$ & 1.9 & 97.4 & 99.3 \\
  $Z=0.0005$, $\rho_g=10^{-8}$, $\dot{m}_3=\dot{m}_{\rm{nuc}}$ & 29.3 & 70.7 & 100 \\
  $Z=0.0005$, $\rho_g=10^{-10}$, $\dot{m}_3=\dot{m}_{\rm{th}}$ & 0.4 & 99.6 & 100 \\ \hline
    \textbf{Model Advanced} \\
  $Z=0.005$, $\rho_g=10^{-8}$, $\dot{m}_3=\dot{m}_{\rm{nuc}}$ & 46.8 & 47.8 & 94.6 \\ 
  $Z=0.005$, $\rho_g=10^{-10}$, $\dot{m}_3=\dot{m}_{\rm{th}}$ & 13.9 & 79.4 & 93.3 \\
  $Z=0.0005$, $\rho_g=10^{-8}$, $\dot{m}_3=\dot{m}_{\rm{nuc}}$ & 26.5 & 73.5 & 100 \\
  $Z=0.0005$, $\rho_g=10^{-10}$, $\dot{m}_3=\dot{m}_{\rm{th}}$ & 0.4 & 99.6 & 100 \\ \hline
    \textbf{Model Basic+NoDrag} \\
  $Z=0.005$, $\dot{m}_3=\dot{m}_{\rm{nuc}}$ & <0.07 & 93 & 93 \\ 
  $Z=0.005$, $\dot{m}_3=\dot{m}_{\rm{th}}$ & <0.07 & 93.1 & 93.1 \\
  $Z=0.0005$, $\dot{m}_3=\dot{m}_{\rm{nuc}}$ & 0.1 & 99.9 & 100 \\
  $Z=0.0005$, $\dot{m}_3=\dot{m}_{\rm{th}}$ & <0.03 & 100 & 100 \\ \hline
    \textbf{Model Advanced+NoDrag} \\
  $Z=0.005$, $\dot{m}_3=\dot{m}_{\rm{nuc}}$ & 19.2 & 67.7 & 86.9 \\ 
  $Z=0.005$, $\dot{m}_3=\dot{m}_{\rm{th}}$ & 6.7 & 79.9 & 86.6 \\
  $Z=0.0005$, $\dot{m}_3=\dot{m}_{\rm{nuc}}$ & 14.4 & 85.6 & 100 \\
  $Z=0.0005$, $\dot{m}_3=\dot{m}_{\rm{th}}$ & 0.4 & 99.6 & 100 \\ \hline
  
\hline\hline                                 
\end{tabular}}
\end{table}

\twocolumn

\begin{table*}[!h]
\caption{Local merger rate of BBHs for systems with TMT investigated in this study.}           
\label{table:merger_rates}      
\centering                         
\resizebox{\textwidth}{!}{\begin{tabular}{l c c c}        
\hline\hline  
& Mergers during TMT ($\rm{Gpc \ yr^{-1}}$) & Mergers after TMT ($\rm{Gpc \ yr^{-1}}$) & Total Mergers ($\rm{Gpc \ yr^{-1}}$) \\
\hline\hline
  \textbf{Model Simple} \\
  $Z=0.005$, $\rho_g=10^{-8}$, $\dot{m}_3=\dot{m}_{\rm{nuc}}$  & 0.56 & <0.01 & 0.56 \\ 
  $Z=0.005$, $\rho_g=10^{-8}$, $\dot{m}_3=\dot{m}_{\rm{th}}$ & 0.56 & <0.01 & 0.56 \\
  $Z=0.005$, $\rho_g=10^{-10}$, $\dot{m}_3=\dot{m}_{\rm{nuc}}$ & <0.01 & 0.56 & 0.56 \\
  $Z=0.005$, $\rho_g=10^{-10}$, $\dot{m}_3=\dot{m}_{\rm{th}}$ & <0.01 & 1.58 & 1.58 \\
  $Z=0.0005$, $\rho_g=10^{-8}$, $\dot{m}_3=\dot{m}_{\rm{nuc}}$ & 0.06 & <0.01 & 0.06 \\
  $Z=0.0005$, $\rho_g=10^{-8}$, $\dot{m}_3=\dot{m}_{\rm{th}}$ & 0.06 & <0.01 & 0.06 \\
  $Z=0.0005$, $\rho_g=10^{-10}$, $\dot{m}_3=\dot{m}_{\rm{nuc}}$ & <0.01 & 0.06 & 0.06 \\
  $Z=0.0005$, $\rho_g=10^{-10}$, $\dot{m}_3=\dot{m}_{\rm{th}}$ & <0.01 & 0.06 & 0.06 \\ \hline
  \textbf{Model Basic} \\
  $Z=0.005$, $\rho_g=10^{-8}$, $\dot{m}_3=\dot{m}_{\rm{nuc}}$  & <0.01 & 0.68 & 0.68 \\ 
  $Z=0.005$, $\rho_g=10^{-8}$, $\dot{m}_3=\dot{m}_{\rm{th}}$ & <0.01 & 0.69 & 0.69 \\
  $Z=0.005$, $\rho_g=10^{-10}$, $\dot{m}_3=\dot{m}_{\rm{nuc}}$ & <0.01 & 0.69 & 0.69 \\
  $Z=0.005$, $\rho_g=10^{-10}$, $\dot{m}_3=\dot{m}_{\rm{th}}$ & <0.01 & 0.89 & 0.89 \\
  $Z=0.0005$, $\rho_g=10^{-8}$, $\dot{m}_3=\dot{m}_{\rm{nuc}}$ & <0.01 & 0.06 & 0.06 \\
  $Z=0.0005$, $\rho_g=10^{-8}$, $\dot{m}_3=\dot{m}_{\rm{th}}$ & <0.01 & 0.06 & 0.06 \\
  $Z=0.0005$, $\rho_g=10^{-10}$, $\dot{m}_3=\dot{m}_{\rm{nuc}}$ & <0.01 & 0.06 & 0.06 \\
  $Z=0.0005$, $\rho_g=10^{-10}$, $\dot{m}_3=\dot{m}_{\rm{th}}$ & <0.01 & 0.06 & 0.06 \\ \hline
  \textbf{Model Basic+BinGDF} \\
  $Z=0.005$, $\rho_g=10^{-8}$, $\dot{m}_3=\dot{m}_{\rm{nuc}}$ & <0.01 & 0.69 & 0.69 \\ 
  $Z=0.005$, $\rho_g=10^{-10}$, $\dot{m}_3=\dot{m}_{\rm{th}}$ & <0.01 & 1.00 & 1.00 \\
  $Z=0.0005$, $\rho_g=10^{-8}$, $\dot{m}_3=\dot{m}_{\rm{nuc}}$ & <0.01 & 0.06 & 0.06 \\
  $Z=0.0005$, $\rho_g=10^{-10}$, $\dot{m}_3=\dot{m}_{\rm{th}}$ & <0.01 & 0.06 & 0.06 \\ \hline
    \textbf{Model Basic+EccGDF} \\
  $Z=0.005$, $\rho_g=10^{-8}$, $\dot{m}_3=\dot{m}_{\rm{nuc}}$  & 0.14 & 0.55 & 0.69 \\ 
  $Z=0.005$, $\rho_g=10^{-10}$, $\dot{m}_3=\dot{m}_{\rm{th}}$ & 0.08 & 0.64 & 0.72 \\
  $Z=0.0005$, $\rho_g=10^{-8}$, $\dot{m}_3=\dot{m}_{\rm{nuc}}$ & <0.01 & 0.06 & 0.06 \\
  $Z=0.0005$, $\rho_g=10^{-10}$, $\dot{m}_3=\dot{m}_{\rm{th}}$ & <0.01 & 0.06 & 0.06 \\ \hline
    \textbf{Model Basic+Iso} \\
  $Z=0.005$, $\rho_g=10^{-8}$, $\dot{m}_3=\dot{m}_{\rm{nuc}}$ & <0.01 & 1.50 & 1.50 \\ 
  $Z=0.005$, $\rho_g=10^{-10}$, $\dot{m}_3=\dot{m}_{\rm{th}}$ & <0.01 & 1.68 & 1.68 \\
  $Z=0.0005$, $\rho_g=10^{-8}$, $\dot{m}_3=\dot{m}_{\rm{nuc}}$ & <0.01 & 0.06 & 0.06 \\
  $Z=0.0005$, $\rho_g=10^{-10}$, $\dot{m}_3=\dot{m}_{\rm{th}}$ & <0.01 & 0.06 & 0.06 \\ \hline
    \textbf{Model Basic+Retro} \\
  $Z=0.005$, $\rho_g=10^{-8}$, $\dot{m}_3=\dot{m}_{\rm{nuc}}$ & 0.23 & 0.40 & 0.63 \\ 
  $Z=0.005$, $\rho_g=10^{-10}$, $\dot{m}_3=\dot{m}_{\rm{th}}$ & 0.02 & 0.70 & 0.72 \\
  $Z=0.0005$, $\rho_g=10^{-8}$, $\dot{m}_3=\dot{m}_{\rm{nuc}}$ & 0.02 & 0.04 & 0.06 \\
  $Z=0.0005$, $\rho_g=10^{-10}$, $\dot{m}_3=\dot{m}_{\rm{th}}$ & <0.01 & 0.06 & 0.06 \\ \hline
    \textbf{Model Advanced} \\
  $Z=0.005$, $\rho_g=10^{-8}$, $\dot{m}_3=\dot{m}_{\rm{nuc}}$ & 0.49 & 0.55 & 1.04 \\ 
  $Z=0.005$, $\rho_g=10^{-10}$, $\dot{m}_3=\dot{m}_{\rm{th}}$ & 0.16 & 1.00 & 1.16 \\
  $Z=0.0005$, $\rho_g=10^{-8}$, $\dot{m}_3=\dot{m}_{\rm{nuc}}$ & 0.02 & 0.04 & 0.06 \\
  $Z=0.0005$, $\rho_g=10^{-10}$, $\dot{m}_3=\dot{m}_{\rm{th}}$ & <0.01 & 0.06 & 0.06 \\ \hline
    \textbf{Model Basic+NoDrag} \\
  $Z=0.005$, $\dot{m}_3=\dot{m}_{\rm{nuc}}$ & <0.01 & 1.02 & 1.02 \\ 
  $Z=0.005$, $\dot{m}_3=\dot{m}_{\rm{th}}$ & <0.01 & 1.01 & 1.01 \\
  $Z=0.0005$, $\dot{m}_3=\dot{m}_{\rm{nuc}}$ & <0.01 & 0.06 & 0.06 \\
  $Z=0.0005$, $\dot{m}_3=\dot{m}_{\rm{th}}$ & <0.01 & 0.06 & 0.06 \\ \hline
    \textbf{Model Advanced+NoDrag} \\
  $Z=0.005$, $\dot{m}_3=\dot{m}_{\rm{nuc}}$ & 0.24 & 1.01 & 1.24 \\ 
  $Z=0.005$, $\dot{m}_3=\dot{m}_{\rm{th}}$ & 0.08 & 1.18 & 1.26 \\
  $Z=0.0005$, $\dot{m}_3=\dot{m}_{\rm{nuc}}$ & 0.01 & 0.05 & 0.06 \\
  $Z=0.0005$, $\dot{m}_3=\dot{m}_{\rm{th}}$ & <0.01 & 0.06 & 0.06 \\ \hline
  
\hline\hline                                 
\end{tabular}}
\end{table*}
\clearpage

\section{Orbit averaged evolution due to gravitational drag}
\label{sec:appA}

In this section, we derive the orbit-averaged derivatives of the semi-major axis and eccentricity for a binary subjected to gravitational drag forces. Built upon the expression for a single orbiting object \citep{murray_solar_1999, grishin_application_2016}, the change in the semi-major axis over time can be described by: 
\begin{equation}
\label{eq:da_dermott}
\frac{da}{dt} = 2\frac{a^{3/2}}{\mu\sqrt{GM(1-e^2)}}[\Delta F_{r}e\sin{\nu}+\Delta F_{\phi}(1+e\cos{\nu})],
\end{equation}
where $\mu = m_1m_2/M$ is the reduced mass, $M=m_1+m_2$ is the total mass of the binary, $\Delta F_{r}$ and $\Delta F_{\phi}$ are the radial and azimuthal components of the gravitational drag force resulting from the differential acceleration between the binary objects, and $\nu$ is the true anomaly of the orbit. The force is given by:
\begin{equation}
    \Delta\vec{F} = \frac{A_{0}m_1^2}{v_1^3}\vec{v}_1 - \frac{A_{0}m_2^2}{v_2^3}\vec{v}_2 = A_0M^2\Bigg(\frac{1}{q^2}+q^2\Bigg)\frac{\vec{v}_b}{v_b^3},
\end{equation}
where $A_0 = 4\pi G^2\rho_g\mathcal{I}$, and the velocities are expressed in terms of the binary velocity: $v_1 = (m_2/M)v_b$ and $v_2 = (m_1/M)v_b$. The minus sign in the expression becomes positive because the velocity vector of the two objects point in opposite direction. The velocity vector in the COM frame is:
\begin{equation}
    \vec{v}_{b} = \frac{na}{\sqrt{1-e^2}}[e\sin{\nu}\hat{r}+(1+e\cos{\nu})\hat{\phi}], 
\end{equation}
where $n = 2\pi/P$ and the velocity magnitude is defined as $v_b = \sqrt{v_{b, r}^2+v_{b, \phi}^2}$. Assuming the gas is at rest with respect to the COM, substituting these expressions into Eq. \ref{eq:da_dermott} yields:
\begin{equation}
\label{eq:da_final}
\begin{split}
    \frac{da}{dt} = \frac{2a^{3/2}}{\mu\sqrt{GM(1-e^2)}}\frac{A_0M^2(1-e^2)}{n^2a^2}\Bigg(\frac{1}{q^2}+q^2\Bigg)\frac{1}{\sqrt{1+2e\cos{\nu}+e^2}} \\
    = \frac{2A_0M^2\sqrt{1-e^2}}{\mu n^3a^2}\Bigg(\frac{1}{q^2}+q^2\Bigg)\frac{1}{\sqrt{1+2e\cos{\nu}+e^2}}.
\end{split}
\end{equation}
Here, we used $\sqrt{GM} = a^{3/2}/n$. To calculate the average derivative over one orbital period, we have:
\begin{equation}
\label{eq:averaged}
    \left\langle\frac{da}{dt}\right\rangle \equiv \frac{1}{P}\int_0^P\frac{da}{dt}dt = \frac{n}{2\pi}\int_0^{2\pi}\frac{da}{dt}\frac{(1-e)^{3/2}}{n(1+e\cos{\nu})^2}d\nu,
\end{equation}
where $dt= d\nu(1-e)^{3/2}/[n(1+e\cos{\nu})^2]$. Combining Eq. \ref{eq:da_final} \& \ref{eq:averaged}, we obtain:
\begin{equation}
\label{eq:app_da}
    \left\langle\frac{da}{dt}\right\rangle = \frac{A_0(1-e^2)^2}{\pi n^3a^2}\frac{M^2}{\mu}\Bigg(\frac{1}{q^2}+q^2\Bigg)\int_0^{2\pi}\frac{dv}{(1+e\cos{\nu})^2\sqrt{1+2e\cos{\nu}+e^2}}.
\end{equation}
For convenience, we kept $\mathcal{I}$ out of the integral, though it depends on the true anomaly. This assumption might influence the accuracy of the calculated values for the semi-major axis derivative. In the case of a circular binary, this expression simplifies to:
\begin{equation}
\begin{split}
    \left\langle\frac{da}{dt}\right\rangle = \frac{2A_0}{n^3a^2}\frac{M^2}{\mu}\Bigg(\frac{1}{q^2}+q^2\Bigg) = \frac{8\pi G^2}{n^3a^2}\rho_g\mathcal{I}\frac{M^2}{\mu}\Bigg(\frac{1}{q^2}+q^2\Bigg) \\ 
    = 8\pi\sqrt{\frac{Ga^5}{M}}\rho_g\mathcal{I}\frac{M}{\mu}\Bigg(\frac{1}{q^2}+q^2\Bigg) \\
    = 8\pi\sqrt{\frac{Ga^5}{M}}\rho_g\mathcal{I}\Bigg[\frac{1}{q}(1+q^{-1})^2 + q(1+q)^2\Bigg].
\end{split}
\end{equation}
The factor between the brackets is similar to the one presented in \citet{rozner_soft_2024}. \newline

Using the same approach, we find the orbit-averaged derivative of the eccentricity. The change in the eccentricity over time is:
\begin{equation}
\frac{de}{dt} =\frac{1}{\mu}\sqrt{\frac{a(1-e^2)}{GM}}[\Delta F_{r}\sin{\nu}+\Delta F_{\phi}(\cos{\nu}+\cos{E})],
\end{equation}
where $E$ is the eccentric anomaly, defined as $\cos{E} = (e+\cos{\nu})/(1+e\cos{\nu})$. Following the same steps as for the semi-major axis, the final expression is:
\begin{equation}
\label{eq:app_de}
    \left\langle\frac{de}{dt}\right\rangle = \frac{A_0(1-e^2)^3}{\pi n^3a^3}\frac{M^2}{\mu}k_q\int_0^{2\pi}\frac{(e+\cos{\nu})dv}{(1+e\cos{\nu})^2(1+2e\cos{\nu}+e^2)^{3/2}},
\end{equation}
where we have defined $k_q = \Bigg(\frac{1}{q^2}+q^2\Bigg)$.
    
\end{appendix}

%
%






   
  



\end{document}